%% file: main.tex
\documentclass[11pt]{article}
\pdfoutput=1 
\usepackage{amssymb, amsmath,amsfonts}
\usepackage{graphicx}
\usepackage{xcolor}
\usepackage{color}
\usepackage[pdftex, bookmarks=true,colorlinks,linkcolor=red,urlcolor=blue,citecolor=blue]{hyperref}
\usepackage[normalem]{ulem}
\usepackage{cite}
\usepackage{subcaption}
\usepackage{slashed}
\usepackage{braket}

\allowdisplaybreaks
\usepackage{enumerate}
\usepackage[normalem]{ulem}

\makeatletter\@addtoreset{equation}{section}\makeatother

\newcommand{\preprint}[1]{\begin{table}[t]  
             \begin{flushright}               
             {#1}                             
             \end{flushright}                 
             \end{table}}                     
\renewcommand{\title}[1]{\vbox{\center\LARGE{#1}}\vspace{5mm}}
\renewcommand{\author}[1]{\vbox{\center#1}\vspace{5mm}}
\renewcommand{\Re}{\operatorname{Re}}
\renewcommand{\Im}{\operatorname{Im}}
\newcommand{\address}[1]{\vbox{\center\em#1}}

\usepackage{bm}
\def\mI{{\bf I}}

\def\be{\begin{eqnarray}}
\def\ee{\end{eqnarray}}
\def\bea{\begin{eqnarray}}
\def\eea{\end{eqnarray}}

\def\Dslash{\,\,{\raise.15ex\hbox{/}\mkern-12mu D}}
\def\Dbarslash{\,\,{\raise.15ex\hbox{/}\mkern-12mu {\bar D}}}
\def\delslash{\,\,{\raise.15ex\hbox{/}\mkern-9mu \partial}}
\def\delbarslash{\,\,{\raise.15ex\hbox{/}\mkern-9mu {\bar\partial}}}
\def\pslash{\,\,{\raise.15ex\hbox{/}\mkern-9mu p}}
\def\calDslash{\,\,{\raise.15ex\hbox{/}\mkern-12mu {\cal D}}}

\newcommand{\Tr}{{\rm Tr}}

\def\lae{\mathrel{\mathop{\smash{\lower .5 ex \hbox{$\stackrel<\sim$}}}}}
\def\lae{\mathrel{\mathop{\smash{\lower .5 ex \hbox{$\stackrel>\sim$}}}}}

\newcommand{\com}[1]{{\bf {\textcolor{cyan}{completed}}}}

\newtheorem{definition}{Definition}[section]

\newtheorem{proposition}{Proposition} [section]

\newtheorem{proof}{Proof}[section]

\begin{document}

\unitlength = .8mm

\begin{titlepage}
\vspace{.5cm}
\preprint{}
\begin{center}
\hfill \\
\hfill \\
\vskip 1cm

\title{\bf Topological invariant for holographic Weyl-$\boldmath Z_2$ semimetal}
\vskip 0.5cm

{Xiantong Chen$^{a}$}\footnote{Email: {\tt chenxiantong23@mails.ucas.ac.cn}}, {Xuanting Ji$^{a,b}$}\footnote{Email: {\tt jixuanting@cau.edu.cn}}, {Ya-Wen Sun$^{a,c}$}\footnote{Email: {\tt yawen.sun@ucas.ac.cn}}

\address{${}^a$ School of Physical Sciences, University of Chinese Academy of Sciences, 
Beijing 100049, China}
\vspace{-10pt}
\address{${}^b$Department of Applied Physics, College of Science, \\
China Agricultural University, Beijing 100083, China}
\vspace{-10pt}
\address{${}^c$Kavli Institute for Theoretical Sciences, \\
University of Chinese Academy of Sciences, 
Beijing 100049, China }
\vspace{-10pt}

\end{center}
\vspace{-10pt}
\abstract{The occurrence of a topological phase transition can be demonstrated by a direct observation of a change in the topological invariant. For holographic topological semimetals, a topological Hamiltonian method needs to be employed to calculate the topological invariants due to the strong coupling nature of the system. We calculate the topological invariants for the holographic Weyl semimetal and the holographic Weyl-$\boldmath Z_2$ semimetal,  which correspond to the chiral charge and the spin-Chern number, respectively. This is achieved by probing fermions within the system and deriving the topological Hamiltonian from the zero-frequency Green's function. In both cases, we have identified an effective band structure characterized by an infinite number of Weyl or $\boldmath Z_2$ nodes, a distinctive feature of holographic systems different from weakly coupled systems. The topological invariants of these nodes are computed numerically and found to be nonzero, thereby confirming the topologically nontrivial nature of these nodes.
}

\setcounter{footnote}{0}

\vfill

\end{titlepage}

\begingroup
\hypersetup{linkcolor=black}
\tableofcontents
\endgroup

\input{Section/Section1}

\input{Section/Section2}

\input{Section/Section3}

\input{Section/Section4}

\input{Section/Section5}
\subsection*{Acknowledgments}
We thank K. Landsteiner and Y. Liu for helpful discussions. This work was supported by the National Natural Science Foundation of China (Grant Nos.12035016, 12405078).
\input{Appendices}

\bibliographystyle{elsarticle-num}
\bibliography{hyperref}

\end{document}

%% file: Section/Section1.tex
\section{Introduction}
\label{sec:1}
Topological phase transition processes represent a novel physical phenomenon that occurs in the topological matter\cite{Witten:2015aoa}. These processes are not encompassed by the Ginzburg-Landau paradigm, which precludes the use of global order parameters for their description\cite{Thouless1, Thouless2, Haldane}. In order to identify the occurrence of a topological phase transition, it is possible to examine the behaviour of specific transport processes, which invariably exhibit distinct characteristics in different topological phases\cite{Landsteiner:2015lsa, Landsteiner:2015pdh, Landsteiner:2016stv, Grignani:2016wyz, Copetti:2016ewq, Bu:2018vcp, Ji:2019pxx, Landsteiner:2019kxb}. In essence, the presence of distinct topological energy band structures gives rise to the varied behaviours observed in transport processes. It hence follows that the topological invariant, which directly indicates the topological structure, is the optimal candidate for distinguishing the various kinds of topological system\cite{Haldane, tknn, km1, moore, km2, qi1}. The topological invariant has been successfully employed in the study of a number of topological systems, including Chern insulators\cite{Haldane} and Weyl semimetals\cite{Armitage, Hasan}, providing valuable insights into their underlying topological properties.

Weyl semimetals represent a novel topological system characterised by the presence of only a few band touching points within the Brillouin zone. These touching points are named as Weyl nodes, which always appear in pairs due to the no-go theorem\cite{Ninomiya, wan}. Accordingly, the fundamental Weyl semimetal is the system that contains only one pair of Weyl nodes, which is referred to as the ideal Weyl semimetal\cite{pan}. The two Weyl nodes exhibit distinct values of the Chern number of $\pm 1$, which represents different chirality. 
From an alternative perspective, the experiments demonstrate that the Weyl nodes can only be annihilated if two Weyl nodes with opposite chiralities merge together.
This result demonstrates the robustness of the Weyl nodes and also confirms the findings of the calculation of the Chern number, which becomes zero when such two Weyl nodes meet together\cite{Armitage}. In light of this, the Chern number can be regarded as a topological invariant of a Weyl semimetal system, which can also be applied to systems that possess additional pairs of Weyl nodes, such as the  Weyl-Z$_2$ semimetal, which features two pairs of Weyl nodes\cite{Ji:2021aan}.

The validity of the aforementioned image depends on the survival of quasi-particles, that is to say, the weak coupling Weyl semimetal.
It is reasonable to inquire as to whether the strong coupling will result in the disruption of the topological structure. Two issues must be addressed in order to examine this further. First, an effective model is required to describe the strong-interacting Weyl semimetal.
Secondly, the difficulty of calculating the Chern number in the strong-coupling regime due to the infinite upper bound of integration presents a challenge. The AdS/CFT correspondence\cite{witten, gubser} and the topological Hamiltonian methods\cite{wang-prx1, wang-prx2, Wang:2012ig} can be employed to address these two issues.
 
AdS/CFT, which is based on the holographic principle of gravity, demonstrates the duality between the classical gravity and a strongly coupled field theory. It represents an effective methodology for the study of strongly coupled many-body systems\cite{Zaanen:2015oix, book, review}. A number of topological semimetals have been the subject of investigation in AdS/CFT, including Dirac semimetals\cite{Bahamondes:2024zsm}, Weyl semimetals\cite{Landsteiner:2015lsa, Landsteiner:2015pdh}, nodal line semimetals\cite{Liu:2018bye, Liu:2020ymx} and  Weyl-Z$_2$ semimetals\cite{Ji:2021aan}. Furthermore, the coexistence of two distinct types of topological semimetals has also been investigated\cite{Chu:2024dti}. These studies contribute to a more profound comprehension of topological systems and gravity.

Another powerful tool is the topological Hamiltonian method, which states that the topological invariants of a strongly coupled system can be calculated from the eigenstates of an effective Hamiltonian, which is the so-called topological Hamiltonian, in a similar manner to that of the weakly coupled theory. The topological Hamiltonian can be derived by calculating the zero-frequency Green's functions, and it exhibits the same topological structure as the original strongly coupled system. It is therefore possible to probe holographic systems using fermions and calculate the corresponding Green's functions. The topological invariant can then be calculated using the topological Hamiltonian. This has been used successfully in the case of holographic Weyl semimetal\cite{Liu:2018djq, Song:2019asj} and the holographic nodal line semimetal\cite{Liu:2018bye, Liu:2020ymx}.

In this work, we study the topological invariants for a different system: the holographic Weyl-Z$_2$ semimetal, which involves two sets of topological invariants: the Weyl charge and the Z$_2$ charge. The calculation of the Weyl charge is similar to the calculation of the topological invariant for the pure holographic Weyl semimetal in \cite{Liu:2018djq}. However, in \cite{Liu:2018djq}, to avoid heavy numerics only analytic calculations for the probe limit were performed, i.e. with no backreaction.  Therefore, in this work, for the study of the more complicated holographic Weyl-Z$_2$ semimetal system, we will first extend the calculation in \cite{Liu:2018djq} for the topological charge to holographic Weyl semimetal with backreaction using numerics. We will first review the procedure for the calculation of the topological invariant for the holographic Weyl semimetal and perform numerical calculations for the topological invariants in parameter regimes where backreaction of matter fields are important, which generalizes the previous probe limit calculation in \cite{Liu:2018djq} to more general circumstances. We examine the fermionic spectral functions of probe fermions on the holographic Weyl semimetal utilizing a numerical method and reveal the existence of multiple Weyl nodes. This is a special feature of holographic systems with backreaction of the matter fields, which does not show up in the probe limit Weyl semimetal calculated in \cite{Liu:2018djq}. We analyze the effective band structures obtained from the topological Hamiltonian and calculate the topological invariants for these nodes using numerics, which turn out to be $\pm 1$, indicating the topologically nontrival nature of the Weyl nodes in holographic Weyl semimetals.

We then proceed to analyze the effective band structures and calculate the topological invariants for the  Weyl-Z$_2$ semimetal utilizing the topological Hamiltonian method. 
The definition of the topological number of a  Weyl-Z$_2$ semimetal differs from that of a typical Weyl semimetal, thus necessitating modification. 
This is because, in order to make clear the topology of a  Weyl-Z$_2$ semimetal which possesses two pairs of nodes, it is not sufficient to consider solely the chirality of the nodes. Given that the degree of freedom of the spin has been introduced in  Weyl-Z$_2$ semimetals, it is logical to conclude that the spin Chern number, a topological invariant that has been used to describe the spin Hall effect in numerous many-body systems\cite{km1}, is the appropriate choice for characterising the topological properties of Weyl-Z$_2$ semimetals.

In light of these considerations, we first undertake a review of the topological band theory, focusing on the physical and geometric significance of the Chern number and facilitating a more intuitive understanding of both the Chern number and the spin-Chern number. We subsequently revisit the effective field theory model\cite{Ji:2023rua} and the holographic model of the Weyl-Z$_2$ semimetal\cite{Ji:2021aan}. The corresponding topological invariant is calculated using the topological Hamiltonian method in the holographic model of the  Weyl-Z$_2$ semimetal and compared with the case in the effective model of weak coupling. Furthermore, the distribution of the Weyl points intrinsic to the holographic Weyl-Z$_2$ semimetal is illustrated through the spectrum of probe fermions. The outcomes demonstrate that the distributions of the aforementioned Weyl points bear both resemblance and difference with those observed in the momentum space of the weak effective field theory model.

The paper is organized as follows. In section \ref{sec:2}, a review of the topological invariant for the Weyl semimetal is provided. New numerical calculations for topological invariants of holographic Weyl semimetal are presented. In section \ref{sec:3}, the effective field theory model and the holographic model of the  Weyl-Z$_2$ semimetal are reviewed, respectively. The effective band structure for the holographic Weyl-Z$_2$ semimetal is obtained through probe fermions, with the requisite zero-frequency Green's functions being derived for the subsequent topological Hamiltonian calculations. In section \ref{sec:4}, the topological Hamiltonian method was employed to calculate the corresponding topological invariant of the holographic  Weyl-Z$_2$ semimetal. Section \ref{sec:5} is devoted to conclusions and open questions.

%% file: Section/Section2.tex
\section{Topological invariant of the Weyl semimetal}
{In this section, we review the basics of the definition and the calculation of the topological invariant in topological states of matter, especially focusing on Weyl semimetals. The usual calculation methods of the topological invariant are based on the weakly coupled band theory, which however does not apply to strongly coupled topological states of matter. Then we review the calculation of the topological invariant of the strongly coupled holographic Weyl semimetal based on the topological Hamiltonian method \cite{wang-prx1, wang-prx2, Wang:2012ig}. 
\label{sec:2}
  \subsection{Basics on topological invariant for Weyl semimetals}

Weyl semimetals are topologically non-trivial phases of matter characterized by the presence of Weyl nodes, i.e. points in momentum space where two non-degenerate bands touch each other, whose low energy excitations are described by the massless solutions of the Weyl equation. These nodes are topologically protected, i.e. they cannot be gapped by small perturbations, such as local perturbations or weak interactions, due to the robustness of the topology for the system. 
To formally characterize this topological property, topological invariants are introduced, which is a mathematical property associated with each Weyl node. In the case of Weyl semimetals, the chiral charge or Weyl charge at each node serves as a key topological invariant. This charge is related to the monopole charge in momentum space, which dictates the topological nature of the nodes. More specifically, the chirality of Weyl nodes is classified by the Chern number, which quantifies the topological winding of the wavefunctions around the nodes. These topological invariants capture the stability of the Weyl nodes against perturbations, characterizing the existence of topological nontrivial Weyl semimetals.

 For topological semimetals, a nonzero Chern number indicates that the energy band of the semimetal cannot be transformed into a trivial band of vacuum, so a nonzero Chern number means that the system is topologically nontrivial and many extraordinary phenomenon may be detected. To define and calculate the topological invariant for a Weyl semimetal, we will introduce a two-dimensional system as an illustrative example to demonstrate the physical significance of the Chern number. Subsequently, we will discuss the definition of the Chern number as it pertains to higher-dimensional systems, which is the situation we will deal with in Weyl semimetal. Finally, we will examine the particular circumstances of a weakly coupled Weyl semimetal, as described by an effective field theory model. Readers who are already familiar with definitions of topological invariants in weakly coupled topological semimetals could skip this part.
 
 In a weakly coupled system, the quasi particle effective Hamiltonian can be used to describe the system nicely. In a many-body system, the electrons are typically assumed to move collectively in accordance with a collective potential. The corresponding eigenstates are the Bloch states, which describe the quasi-continuous movement of the significant number of electrons in the momentum space. The eigenvalues of the Bloch states are therefore constituted as a quasi-continuous structure, which is designated an ``energy band." The following example will illustrate the application of energy band theory to a two-dimensional topological system. 
            
            Consider a weak coupling system which is described by a Hamiltonian $\hat{H}(k)$, at every $k$ in momentum space, the eigenvalue $E_{n}(k)$ and corresponding eigenvector $\ket{n(k)}$ can be obtained by solving the eigen-function of $\hat{H}(k)$. Take only one branch of eigenvalue $E_n(k)$ and the corresponding eigenvector $\ket{n(k)}$ as an example, we have 
            \begin{equation}
            \label{eigen}
            \hat{H}(k)\ket{n(k)}=E_n(k)\ket{n(k)}  .
            \end{equation}
            
            The corresponding Berry connection $\omega$ and Berry curvature $\Omega$ can be calculated using $\ket{n(k)}$
            \be
            \label{BerryConnectionCurvature}
                \omega=i\braket{n(k)|d|n(k)},~\Omega=id\braket{n(k)|\wedge d|n(k)}.
            \ee
            
            Then the Chern class $c_1$ for the energy band $E_n$ is defined as 
            \begin{equation}
                \label{ChernClass}
               c_1(E_n)=-\frac{[\Omega]}{2\pi i}=-\frac{[d\braket{n(k)|\wedge d|n(k)}]}{2\pi}.
            \end{equation}
            
            Notice that $[\Omega]$ is the class which Berry curvature $\Omega$ represents in de Rham cohomology group, i.e. for an arbitrary exact 2-form $\eta$, $[\Omega]=[\Omega+\eta]$. When we need to integrate $[\Omega]$ over a closed surface, we just integrate Berry curvature $\Omega$ over the surface, for the integration of any exact form over a closed surface is $0$.  Now given a 2-dimensional closed surface $M$ in momentum space, we can integrate the Chern class $c_1$ all over $M$ to get the Chern number $C$ corresponding to energy band $E_n$ 
            \begin{equation}
                \label{CharacteristicNumber}
                C(E_n)=\int_{M}c_1=-\int_M\frac{d\braket{n(k)|\wedge d|n(k)}}{2\pi}.
            \end{equation}
            \subsection{Topological invariant in effective field theoretical model for Weyl semimetal}
            \label{TopologicalInvariantInPureAdS}

           Having reviewed the theory of energy bands and the definition of the Chern number as the topological invariant, we will apply them in the case of the Weyl semimetal and present an effective field theory model of the ideal Weyl semimetal, along with a definition of the Weyl charge.

            Topological semimetals can be described by effective models that generalise intuitively\textcolor{purple}{?} from the relativistic field theory model with Lorentz symmetry breaking\cite{Grushin:2012mt, Grushin:2019uuu}. In the case of the Weyl semi-metal, the corresponding effective field theory model can be derived from the Dirac equation, as the two Weyl points with opposite chirality can be obtained by breaking either time-reversal or inversion symmetry in a Dirac point. Consider the Dirac equation with mass $m$,
            \begin{equation}
               i\gamma^\mu\partial_\mu\psi=m\psi, 
            \end{equation}
            after adding a ``Lorentz breaking" term in the equation, we obtian
            \begin{equation}
            (i\gamma^\mu(\partial_\mu+iA_\mu)-b_\mu\gamma^5\gamma^\mu)\psi=m\psi,
            \end{equation}
            where $A_\mu$ is electromagnetic  gauge field and $b_\mu$ is the axial gauge field. The action of the ideal Weyl semimetal can be written as follows
            \begin{equation}
                \label{idealaction}
                S=\int d^4x\Bar{\psi}(i\gamma^\mu\partial_\mu-b_\mu\gamma^5\gamma^\mu-m)\psi,
            \end{equation}
            and the corresponding Hamiltonian is
            \begin{equation}
                \label{idealHamiltonian}
                H=\int d^3x\psi^{\dag}(-i\gamma^0\gamma^i\partial_i+b_\mu\gamma^0\gamma^5\gamma^\mu+m\gamma^0)\psi.
            \end{equation}
            
            The term $b_0\gamma^5\gamma^0$ breaks the parity symmetry and $b_i\gamma^5\gamma^i$ is the time-reversal breaking term. Without loss of generality we choose $b=b_z \delta_{i}^{z}$ and observe the Fourier transformation of Hamiltonian. We now consider the case in the momentum space, the corresponidng Hamiltonian can be written as:
            \begin{equation}
            \label{WeylHamilton}
                \hat{H}(k)=\begin{pmatrix}
                    -k_i\sigma^i-b_z\sigma^3 & m\\
                    m & k_i\sigma^i-b_z\sigma^3
                \end{pmatrix},
            \end{equation}
            with the corresponding dispersion relation when $k_x=k_y=0$:
            \begin{equation}
                \label{idealspectrum}
            \omega=\pm(b_z\pm\sqrt{k_z^2+m^2}).
            \end{equation}
            
             \begin{figure}[h]
         \begin{minipage}{0.475\linewidth}
                \vspace{3pt}
                \centerline{\includegraphics[width=\textwidth]{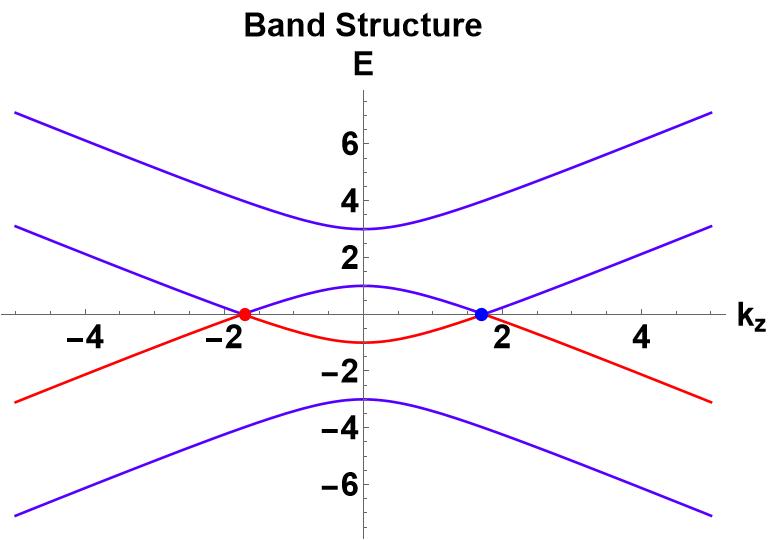}}
            \end{minipage}
            \begin{minipage}{0.5\linewidth}
                \vspace{3pt}
                \centerline{\includegraphics[width=\textwidth]{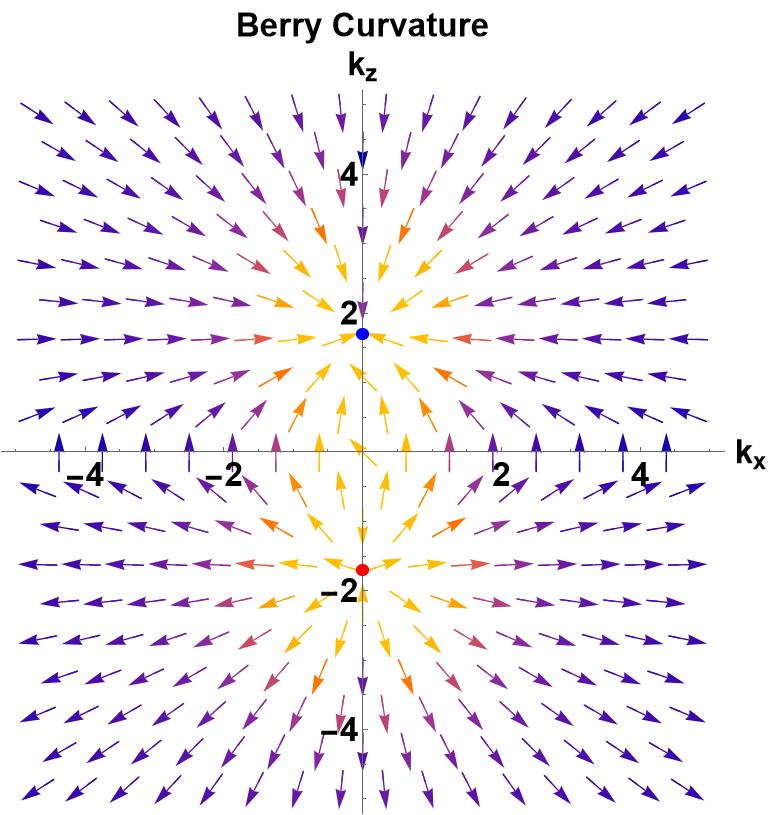}}
            \end{minipage}
            \caption{The band structure of \eqref{WeylHamilton} along $k_z$ axis and the corresponding Berry curvature, which has been Hodge star operated, on $k_x-k_z$ plane. There are two Weyl nodes which has been marked red for Weyl charge +1, blue for -1. Without loss of generality we choose $b=2,m=1$ as an example in the system.}
            \label{fig1}
        \end{figure}
            
            The left panel of Fig. \ref{fig1} illustrates the eigenenergy, otherwise termed the dispersion relation \eqref{idealspectrum}, of the Hamiltonian \eqref{WeylHamilton} in relation to $k_z$. Without loss of generality, it is assumed that $b_z=2$ and $m=1$ in \eqref{idealspectrum}. Two band crossing points are formed and located at $k_z=\pm\sqrt{3}$, respectively. It can be posited that the aforementioned two crossing points can be regarded as two Weyl nodes, which can in principle be obtained from a single Dirac point. The low-energy excitation in the vicinity of the Weyl node can be described using the Weyl equation, which implies that the elementary excitation is a Weyl fermion. Each of the nodes possesses a unique chirality, which can be distinguished by the topological number, as can be obtained by integrating the Berry curvature, \eqref{ChernClass}, all over a closed surface. As illustrated in the right-hand panel of Fig. \ref{fig1}, the corresponding Berry curvature is demonstrated. It can be observed that the two Weyl nodes act as a sink and a source for the Berry curvature, respectively, thus indicating that these two nodes possess opposite values of the topological number. The result indicates that the topological number for the red point is $+1$, while the blue point possesses a negative topological number. The classification of two Weyl nodes is dependent upon their topological number. 

            In summary, in a weak coupling situation, the corresponding Chern class, designated as $c_1$, can be calculated once a branch of eigenvectors has been obtained. The integration of $c_1$ over a closed surface $M$ allows for the identification of the presence of a Weyl point within the surface $M$. 
            
            The valence band is not occupied. In the numerical calculation, we usually use the following foumula to calculate the Chern class of the system, i.e. $c_1$ is the sum of the Chern class for filled bands, which is much more stable for numerical calculation.
            \be
        c_1=\left[\sum_{E_n\le E_F}-\frac{d\bra{n(k)}\wedge d\ket{n(k)}}{2\pi}\right],
        \ee
        where $E_F$ is the Fermi energy.
        
        As a result, the topological invariant, can be designated as a Weyl charge, which bears resemblance to electric charge. The Weyl charge surrouneded by closed surface $M$ can be defined as
                \be
                Q:=\int_M c_1,
                \ee
        
\subsection{Topological invariant in holographic Weyl semimetal: the topological Hamiltonian method}
The calculation of the topological number can be achieved through the surface integral, a process which may prove challenging to integrate out. {As there is no Hamiltonian for a strong coupling system,} with the aid of the Green's function, the topological number can also be calculated as follows:
    \be
    \label{another}
    N(k_z)=\frac{1}{24\pi^2}\int dk_0 dk_x d k_y \Tr\Big[\epsilon^{\mu\nu\rho z}G\partial_{\mu}G^{-1}G\partial_{\nu}G^{-1}G\partial_{\rho}G^{-1}\Big]\,,
    \ee
where $\mu,\nu,\rho \in k_0,k_x,k_y$ and $k_0=i\omega$ is the Matsubara frequency. It has been demonstrated that this integral is applicable to both non-interacting and interacting systems. However, when considering interactions, integration in the $k_0$ direction becomes inevitable, and is challenging to calculate in the strong coupling limit. Notwithstanding the fact that the  Green's function for any value of the variable $\omega$ can be obtained numerically in holography, this process is extremely time consuming and numerical inaccuracies may also be accumulated.

In order to circumvent the aforementioned limitations, an effective Hamiltonian need to be introduced for the system. This is to be defined by the zero frequency Green's function, and it has been shown to comprise all the topology information of the system. This has been termed the ``topological Hamiltonian"\cite{wang-prx1, wang-prx2, Wang:2012ig}. In a sense, the zero frequency Green's function $G(0,k)$ contains all the information of $G(\omega,k)$ for $\omega$ takes an arbitrary value. All the poles of $G(\omega,k)$ are located on the $\omega=0$ plane, just like that a meromorphic function whose poles all locates at real axis. It is widely known that the pole contains all the information of the function, i.e. the near real axis part of the function can confirm the whole function. In this subsection, the definition of the topological Hamiltonian is provided. Subsequently, the ideal Weyl semimetal case is employed as a demonstrative exemplar to illustrate the calculation of the topological invariant in the holographic model by the topological Hamiltonian.   
        \subsubsection{The topological Hamiltonian method}
            
            {As has been introduced in the previous part, in order to calculate to topological invariant in strong coupling systems where we do not have a known working Hamiltonian, we have to deal with the integral along $\omega$ axis utilizing the formula \eqref{another}, which is a very time consuming procedure. However, there is a more efficient method, namely the topological Hamiltonian method. The basic idea is that though the quasiparticle effective Hamiltonian is invalid, the zero frequency Green's function $G(0,k)$ contains all the information of topological properties of $G(\omega,k)$. Thus we can use the zero frequency Green's function $G(0,k)$ to define a so-called ``topological Hamiltonian" $H_{eff}$. Then we could use this effective Hamiltonian to calculate the topological charge as if we are in a weak coupling system.}
            
            Given the zero frequency Green's function $G(0,k)$ of a system, the topological Hamiltonian can be defined as
              \be
              \label{toh}
              H_{eff}(k):=-G^{-1}(0,k).
              \ee
             
            Notice that if we choose $\omega$ to be exactly zero, there will be some $k$ where $G(\omega,k)$ is not reversible, so we always choose $\omega$ as a very small value such as $10^{-12}$ and use $H_{eff}(k)=-G^{-1}(\omega,k)$. 
                
                Although the topological Hamiltonian is not the exact Hamiltonian of the strong coupling system (there is not a Hamiltonian that can be used to describe the whole strong coupling system), the topological Hamiltonian can be used to give the exact topological invariant, which cannot be calculated from the  quasiparticle effective Hamiltonian. In conclusion, as long as ther Green's function $G(\omega,k)$ does not have any poles besides on the $\omega=0$ plane, we can treat the topological Hamiltonian as a Hamiltonian in weak coupling system to calculate the topological invariant as it has been stated in previous works\cite{Wang_2013}.{\footnote{Note that in the original proposal \cite{wang-prx1, wang-prx2, Wang:2012ig}, for the topological Hamiltonian method to work, the regularity of the Green functions in the imaginary frequency space has to be required. Here because we have access to the more physical real frequency Green functions, we instead use the constraint that the real frequency Green functions are regular near the semimetal nodes, which are easy to be confirmed. Detailed analysis on this point is in Appendix \ref{AppendixD}.}}
                
    \subsubsection{Review of the holographic Weyl semimetal}
        In a strongly coupled system, the description of the low-energy excitation of the quasi-particle, as well as the perturbation theory, is not applicable. Furthermore, the symmetries of some systems are also violated. Consequently, it is challenging to utilise an explicit Hamiltonian to investigate a strongly coupled many-body system. However, the gauge/gravity duality allows for the construction of a holographic model to describe the many-body system in the strong coupling limit. Subsequently, the system can be solved using weak classical gravity. The 
       bulk action for the holographic Weyl semimetal is\cite{Landsteiner:2015lsa, Landsteiner:2015pdh}.
         \be
        \label{holoideal}
          S&=&\int d^5x \sqrt{-g}\bigg[\frac{1}{2\kappa^2}\Big(R+\frac{12}{L^2}\Big)-\frac{1}{4}{F_V}^2-\frac{1}{4}{F_A}^2+\frac{\alpha}{3}\epsilon^{abcde}A_a(3{F_V}_{bc}{F_V}_{de}+{F_A}_{bc}{F_A}_{de}) \nonumber\\
&~~~ -&(D_a\Phi)^*(D_a\Phi)-m|\Phi|^2-\frac{\lambda}{2}|\Phi|^4\bigg]\,,
\ee
        where $V$ is a vector $U(1)$ gauge field and $A$ is the axial $U(1)$ gauge field. $F_V,F_A$ are the corresponding field strength, respectively. The coefficient $\alpha$ corresponds to the Chern-Simons term and is associated with the chiral anomaly. The scalar field $\Phi$ is charged only under the axial gauge symmetry. The corresponding covariant derivative is defined as $D_a:=\partial_a-i q A_a$. $R$ is the scalar curvature. The scalar bulk mass is set to be $m^2=-3$, and the radius of the AdS $L=1$ without loss of generality. We will choose other parameters in \eqref{holoideal} to be  $2\kappa^2=1,\alpha=1,\lambda=\frac{1}{10}$ and $q=1$ without loss of generality. 
        
        At zero temperature, the solution can be parametrized as
        \be
        \label{idealansatz}
        ds^2=u(r)(-dt^2+dx^2+dy^2)+h(r)dz^2+\frac{dr^2}{u(r)},~\Phi=\phi(r),~A=A_z(r)dz,~V=0,
        \ee  
        where $u,h,\phi,A_z$ are all real functions of the radial coordinate $r$. In the vicinity of the asymptotic AdS boundary, the leading order asymptotic behavior of these functions should be
        \begin{equation}
           u=r^2+\cdots,~h=r^2+\cdots,~\phi=\frac{M}{r}+\cdots,~A_z=b+\cdots. 
        \end{equation}
        
        \subsubsection{Green's functions from probe fermions}
            As previously stated, the topological invariant of the ideal Weyl semimetal in the weak coupling limit can be calculated by integrating the $c_1$ Chern class over a closed surface. However, in the strong coupling case, we do not have information on the eigenstates of the system so that the direct calculation of the topological invariant cannot be used anymore. If we use the formula \eqref{another} to calculate the topological invariant, another problem arises as
            the upper bound of the integration along the omega axis becomes infinite,  which complicates the calculation of the integral. In the strong coupling cases, we will utilize  the topological Hamiltonian method, which has been introduced above to be valid in this case. The topological Hamiltonian is defined as the zero-frequency Green's function so that eigenstates could be well defined to be used to calculate the topological invariant. Consequently, the topological Hamiltonian is a powerful tool for calculating the topological invariant in the strong coupling regime.

            In the holographic model, it is possible to probe the system with fermions and thereby obtain the corresponding zero-frequency Green's function. The effective topological Hamiltonian is then determined, and the corresponding topological invariants can be calculated in a straightforward manner. The efficacy of this method has been demonstrated in a holographic ideal Weyl semimetal\cite{Liu:2018djq} and a topological nodal-line semimetal\cite{Liu:2018bye, Liu:2020ymx}.

            In the following, we will first introduce the basics of probe fermions in the holographic Weyl semimetal and review the topological Hamiltonian method in this holographic framework. 
    In the study of Weyl semimetal, fermions with both chiralities are important in forming the Weyl nodes, so we need to introduce holographic fermions which correspond to both two chiralities, which is different from the framework of only one chirality for probe fermions previously used in the RN black hole background \cite{Cubrovic:2009ye ,Liu:2009dm}. To describe probe fermions that correspond to both chiralities on the boundary, we need to introduce two probe ferimons in the bulk, each coupling to the bulk $A_\mu$ and $\Phi$ fields in a different way to produce opposite chiralities \cite{Liu:2018djq}. The coupling to $\Phi$ produces mass for the fermions on the boundary \cite{Plantz:2018tqf}.
            
              The procedures are enumerated as follows: firstly, the action of the probe fermions will be introduced; following this, the definition of the Green's function will be provided. We will use a pair of fermions $\Psi_1,\Psi_2$ with opposite mass to detect the chirality of the system and get the Green's function. Subsequently, an illustration of this definition in the case of pure AdS will be presented to demonstrate the validity of the definition; and finally, topological Hamiltonian method for holographic Weyl semimetal, {which aims to calculate the Weyl charge} based on the Green's function will be presented in the next subsection.  The implementation of these procedures enables the calculation of the corresponding topological invariant in a holographic model, as in a weakly coupled system.
            
            {Our discussion starts from the action of the probe fermions}
        \be
        \label{probeideal}
            S_{fermion}&=&\int d^5x\sqrt{-\det g}\Bigg[\Psi_1(\Gamma^aD_a-m-i A_a\Gamma^a)\Psi_1+\Psi_2(\Gamma^aD_a+m+i A_a\Gamma^a)\Psi_2 \nonumber \\
           &-&(\eta\Phi\Bar{\Psi}_1\Psi_2+\eta^*\Phi^*\Bar{\Psi}_2\Psi_1)\Bigg],
        \ee
        where the wave functions of the two probe fermions are denoted by $\Psi_{1,2}$, respectively. The coupling strength between the probe fermions and the scalar field $\Phi$, in the holographic model \eqref{holoideal} is described by the parameter, denoted by $\eta$. This parameter can be set to $\eta=1$ without loss of generality. $m$ denotes the mass term of probe fermions. In this work, it has been set to $\frac{1}{4}$. The corresponding gamma matrix is denoted by $\Gamma^a$, and the corresponding covariant derivative by $D_a$. The definitions of these can be found in appendix \ref{AppendixA}.         
        
        To calculate the holographic Green's function, we will need to calculate the wave function of probe fermions in the holographic Weyl semimetal background and the boundary terms of the action. Here before the calculation on the Weyl semimetal background, we will start from a review of the probe fermions on the pure AdS background, in which the procedure will be made clear and the asymptotic boundary behavior will be straight forward to analyze. This simple illustration will help pave the background for the calculation of the probe fermions in the holographic Weyl semimetal background in the next subsection. 
        
            For pure AdS spacetime, the field $A_z=0,\phi=0$, and the metric has components $u(r)=h(r)=r^2$. Thus the equation of motion for each fermion has the form
               \be r\Gamma^r\frac{\partial}{\partial r}\psi_{1,2}+\frac{ik_\mu\Gamma^\mu}{r}\psi_{1,2}- m_{1,2}\psi_{1,2}=0,
            \ee where $\Psi_{1,2}=\psi_{1,2}(r)e^{ik_\mu x^\mu}$  with $m_1=-m_2=m$. The two probe fermions do not couple with each other, so we only need to discuss the equation of motion for either one fermion. Without loss of generality we take $\psi=\psi_1$ as an example to solve it under the infalling boundary condition. The solution of $\psi_2$ can be solved using the same method. 
            
            We apply $\Gamma_\pm:=\frac{1\pm \Gamma^r}{2}$ on the two sides of the equation, which projects the spinor to the chiral left mode $\psi_+$ and right mode $\psi_{-}$. The equation can be simplified as
                \be r\frac{\partial}{\partial r}\psi_{+}+\frac{ik_\mu\Gamma^\mu}{r}\psi_{-}- m\psi_{+}&=&0,\nonumber\\
                -r\frac{\partial}{\partial r}\psi_{-}+\frac{ik_\mu\Gamma^\mu}{r}\psi_{+}- m\psi_{+}&=&0,
                \ee
              which can be written into the following form
              \be \psi_{-}&=&\frac{ik_\mu{\Gamma}^\mu}{k^2}A(m)\psi_{+},\nonumber\\
              \psi_{+}&=&-\frac{ik_\mu{\Gamma}^\mu}{k^2}A(-m)\psi_{-},
              \ee 
                with $A(m):=r\left(r\frac{\partial}{\partial r}-m\right)$. Then the chiral left and right modes $\psi_{\pm}$ can be separated as
                \be k^2\psi_{+}&=&A(-m)A(m)\psi_{+},\nonumber\\
                k^2\psi_{-}&=&A(m)A(-m)\psi_{-}.
                \ee
                Taking the first equation $k^2\mathfrak{F}=A(-m)A(m)\mathfrak{F}$ as an example,  we could perform the transformation  $\mathfrak{F}=\frac{F}{\sqrt{r}},~r=\frac{k}{z}$ and the equation becomes a Bessel equation whose solution is known
            \be
            z^2\frac{d^2 F}{dz^2}+z\frac{dF}{dz}-\left(z^2+\left(m+\frac{1}{2}\right)^2\right)F=0,
             \ee
             \be \mathfrak{F}=\frac{C_1}{\sqrt{r}}I_{m+\frac{1}{2}}\left(\frac{k}{r}\right)+\frac{C_2}{\sqrt{r}}K_{m+\frac{1}{2}}\left(\frac{k}{r}\right).
             \ee
            
                Imposing the boundary condition that $\mathfrak{F}$ is finite at $r=0$, we only reserve the Bessel $K$ function as a solution, i.e. the physical solution for probe fermions on the pure AdS geometry is
               \be \psi_{+}&=&\frac{1}{\sqrt{r}}K_{m+\frac{1}{2}}\left(\frac{k}{r}\right)a_+,\nonumber\\
                \psi_{-}&=&\frac{ik_\mu\Gamma^\mu}{k^2}A(m)\psi_{+}=\frac{ik_\mu\Gamma^\mu}{k}\frac{1}{\sqrt{r}}K_{m-\frac{1}{2}}\left(\frac{k}{r}\right)a_+,~~~~a_+\in \Im\Gamma_+.
                \ee
                
                Then using the series definition of Bessel K function, we can investigate the behaviour of the solution near the boundary
                \be \psi_{+}&=&\frac{k^{-m-\frac{1}{2}}}{2^{\frac{1}{2}-m}}\Gamma\left(m+\frac{1}{2}\right)r^m a_+,\nonumber\\
                \psi_{-}&=&\frac{k^{m-\frac{1}{2}}}{2^{\frac{1}{2}+m}}\Gamma\left(-m+\frac{1}{2}\right)\frac{1}{r^m}\frac{ik_\mu\Gamma^\mu}{k} a_+,~~~~r\to\infty.
                \ee
                
                We can observe that $\psi_+$ goes to infinity as $r\to\infty$, while $\psi_-$ stays finite, so it is reasonable to regard $\psi_+$ as an source and $\psi_-$ is a response which is fully determined by the source term. We could perform the same procedure for the fermion with negative mass to get the corresponding solutions.
                
%
%

            {After discussing the solution of probe fermions in pure AdS background, we will investigate the boundary term of the action, which determines the structure for Green's function in detail. To do this, we start from the bulk action $S_{bulk}=\int d^{d+1}x \sqrt{-g}(\Bar{\psi}\Gamma^a D_a\psi-m\Bar{\psi}\psi)$ term in the $d+1$ dimensional AdS space, and let us observe its variation on the d-dimensional boundary}
            \be
            \delta S_{bulk}=\int\sqrt{-h}d^{d}x(\Bar{\psi}_-\delta\psi_+-\Bar{\psi}_+\delta\psi_-), 
            \ee where $h$ is induced metric of $g$ on the boundary.
            
            However, as we have pointed out that $\psi_-$ is the response of $\psi_+$, the term $\delta\psi_-$ should not appear in the boundary variation. Thus we have to add a boundary term to make it vanish
            \be S_{boundary}=\int d^d x\sqrt{-h}\Bar{\psi}_{+}\psi_-.
                \ee
            
            {As a result, we can claim that in d+1 dimensional AdS space, the action for a probe fermion can be written as the sum of two parts, i.e. $S=S_{bulk}+S_{boundary}$, where}
             \be 
             S_{bulk}=\int d^{d+1}x\sqrt{-g}(\Bar{\psi}\Gamma^a D_a\psi-m\Bar{\psi}\psi),~S_{boundary}=\int d^d x\sqrt{-h}\Bar{\psi}_{+}\psi_-.
                \ee

            In the case of two probe fermions with opposite masses, {as $\Psi_{1-}$ and $\Psi_{2+}$ are the response terms and $\Psi_{1+}$ and $\Psi_{2-}$ are the source terms}, the corresponding boundary terms can be expressed as follows  :
            \be
            S_{1,boundary}&=&\int d^d x\sqrt{-h}\Bar{\Psi}_{1+}\Psi_{1-},\nonumber\\
            ~S_{2,boundary}&=&-\int d^d x\sqrt{-h}\Bar{\Psi}_{2-}\Psi_{2+},
            \ee
           therefore, it is apparent that the overall boundary term for the system involving two probe fermions with opposite masses is following:
          \be
          \label{idealboundary}
          S_{boundary}=\int d^d x\sqrt{-h}(\Bar{\Psi}_{1+}\Psi_{1-}-\Bar{\Psi}_{2-}\Psi_{2+}).
          \ee
        It can be demonstrated that the boundary term \eqref{idealboundary} can be expressed as follows:
        \be
         S_{boundary}=\int d^d x\sqrt{-h}\Bar{\Psi}\chi=\int d^d x\sqrt{-h}\Psi^\dag\Gamma^0\chi,
         \ee
         wherein the following definitions are employed:
         $\Psi:=\Gamma_+\Psi_1+\Gamma_-\Psi_2,~\chi:=\Gamma_-\Psi_1-\Gamma_+\Psi_2$. The Green's function can be defined with these two {combined fields $\Psi$ and $\chi$.} {$\Psi$ is the source and $\chi$ is the response term, the corresponding relationship between which can be written as $\chi=-i\Xi\Psi$, and} the Green's function is
               \be G_{chirality}:=\lim_{r\to\infty}r^{2m}\Gamma^0\Xi.\ee
          {Note that in pure AdS, the behavior of the two chiralities of fermions have no difference due to vanishing $A_z$ and $\Phi$, but we will see in the next subsection that different chiralities are important in obtaining the correct band structure of the holographic Weyl semimetal.}

        \subsubsection{Basic procedure for topological invariant of holographic Weyl semimetal}
            In the following, the calculation of the topological invariant through the topological Hamiltonian method in the holographic Weyl semimetal will be demonstrated. In order to facilitate this process, it is essential to employ an effective Hamiltonian, which is defined by the zero-frequency Green's function and is known as the topological Hamiltonian\cite{wang-prx1, wang-prx2, Wang:2012ig}. 
            In order to calculate the Green's function for the holographic Weyl semimetal, it is necessary to solve the equations of motion for the probe fermions, which can be obtained from \eqref{probeideal}. The probe fermion fields can be expanded as  
            \be
             \Psi_l=\psi_l e^{-i\omega t+i k_x x+i k_y y+i k_z z} \,,~~~~l=1,2\,.
             \ee

            {As previously mentioned, the introduction of two probe fermions is indicative of the presence of two chiralities, which are present at the boundary field theory, i.e. the chirality observed in the Weyl semimetal. 
             The boundary chirality is manifested in the bulk through the manner in which fermions couple with the vector field (i.e. $A_z$) and the scalar field (i.e. $\Phi$), as illustrated in \eqref{probeideal}.}

            {The equations of motion for the probe fermions can then be obtained by substituting the background geometry \eqref{idealansatz} of the holographic Weyl semimetal:}
           \be
           \label{eom1}
        &\sqrt{u}\Gamma^{{r}}\partial_r\psi_1+\left(-i\frac{\omega}{\sqrt{u}}\Gamma^{{t}}+i\frac{k_x}{\sqrt{u}}\Gamma^{{x}}+i\frac{k_y}{\sqrt{u}}\Gamma^{{y}}+i\frac{(k_z-q A_z)}{\sqrt{h}}\Gamma^{{z}}\right)\psi_1-m_f\psi_1-\eta\phi\psi_2=0,\nonumber\\
           & \sqrt{u}\Gamma^{{r}}\partial_r\psi_2+\left(-i\frac{\omega}{\sqrt{u}}\Gamma^{{t}}+i\frac{k_x}{\sqrt{u}}\Gamma^{{x}}+i\frac{k_y}{\sqrt{u}}\Gamma^{{y}}+i\frac{(k_z+q A_z)}{\sqrt{h}}\Gamma^{{z}}\right)\psi_2+m_f\psi_2-\eta\phi\psi_1=0.
            \ee
           
           In the holographic Weyl semimetal, the topological nontrivial phase is characterised by the following near-horizon geometry   
           \be
           \label{BackGroundIRGeometry}
           u=r^2,~h=r^2,~A_z=a_1+\frac{\pi a_1^2\phi_1^2}{16 r}e^{-\frac{2 a_1}{r}},~\phi=\sqrt{\pi}\phi_1\left(\frac{a_1}{2r}\right)^{\frac{3}{2}}e^{-\frac{a_1}{r}}.
           \ee
           It is noteworthy that the geometry above demonstrates that, in the limit of minimal $r$, it can be deduced that $u$ and $h$ are equivalent to $r^2$, and $A_z$ approximates $a_1$, with $\phi$ taking on a value of approximately 0. This configuration bears a striking resemblance to the pure AdS case. Consequently, the wave function of the probe fermion near the horizon can be expanded as follows
           \be
           \label{NearHorizon}
           {\psi_{1}}_{horizon}&=\frac{1}{\sqrt{r}}K_{m+\frac{1}{2}}\left(\frac{k_1}{r}\right)a_+ +\frac{ik_{1\mu}\Gamma^\mu}{k_1}\frac{1}{\sqrt{r}}K_{m-\frac{1}{2}}\left(\frac{k_1}{r}\right)a_++\cdots,~a_+\in \Im\Gamma_+,\nonumber\\
            {\psi_{2}}_{horizon}&=\frac{1}{\sqrt{r}}K_{m+\frac{1}{2}}\left(\frac{k_2}{r}\right)a_- +\frac{ik_{2\mu}\Gamma^\mu}{k_2}\frac{1}{\sqrt{r}}K_{m-\frac{1}{2}}\left(\frac{k_2}{r}\right)a_-+\cdots,~a_-\in \Im\Gamma_-,
            \ee
            where $k_l:=(-\omega,k_x,k_y,k_z+(-1)^l a_1),~l=1,2$, {and $\cdots$ denotes higher order terms}.

            In the case of $r \to \infty$, that is, in the vicinity of the boundary of the AdS, the leading order  behavior of the probe fermions is
            \be
            \label{UVbehaviour}
            \psi_1\stackrel{r\to\infty}{\longrightarrow}\begin{pmatrix}{a_1}^1r^{m_f}\\ {a_1}^2r^{m_f}\\ {a_1}^3r^{-m_f}\\ {a_1}^4r^{-m_f} \end{pmatrix},~\psi_2\stackrel{r\to\infty}{\longrightarrow}\begin{pmatrix}{a_2}^1r^{-m_f}\\ {a_2}^2r^{-m_f}\\ {a_2}^3r^{m_f}\\ {a_2}^4r^{m_f}\end{pmatrix}.
            \ee

            In light of the aforementioned behaviour exhibited by the probe fermions, the retarded Green's function can be calculated directly. As previously established, the equations in \eqref{eom1} are comprised of two coupled first-order Dirac equations. When the in-falling boundary conditions are applied at the horizon, it is possible to select four different initial conditions at the horizon. The subsequent solution of these equations results in near boundary values, thereby determining the retarded Green's functions.
            
            {The calculation of the holographic Green's function and the consequent topological invariant for the holographic Weyl semimetal in the limit $M/b \to 0$, which corresponds to the case where no back reaction of $A_z$ was considered}, and for small $M/b$, i.e. $\frac{M}{b} \ll\frac{M}{b}_{(c)}$, which corresponds to a small back reaction}, have been obtained in \cite{Liu:2018djq} using an analytic or semi-analytic method. However, for general $\frac{M}{b}$, where the backreaction becomes large, the analytical method becomes invalid. If backreaction is not taken into account, the system would be in the $M/b \to 0$ limit, then several interesting features of the effective band structure would be lost: 
\begin{itemize}
    \item There will only be one pair of Weyl nodes as in Fig.4, which is much similar to that in weak coupling system.
    \item The two gapped bands would seem to have a crossing as in [36], which is an artifact of the $M/b \to 0$ limit.
    \item Multiple Weyl nodes and the band inversion behavior - features that will be shown to exist in the backreaction case - would be lost.
    \end{itemize}

           Therefore, the present work employs numerical methods to facilitate the calculation of backreaction cases and the results of this calculation are presented in subsection \ref{wfs}. The calculation process is generally comprised of the following three steps.

           The initial step in this process is to produce the numerical background for the holographic Weyl semimetal.
         The second step is to calculate the solutions of probe fermions on the holographic Weyl semimetal background numerically. We pick a very small $r_0>0$ and set the in-falling boundary condition for $\psi_{horizon}$. Before proceeding, it is necessary to clarify the physical meaning of $r_0$. The Lifshitz horizon in \eqref{BackGroundIRGeometry} is situated at the point of $r=0$. However, in order to solve the equations of motion numerically, it is necessary to employ an IR cutoff, designated as $r_0$, in order to import the initial condition of the Lifshitz IR geometry, i.e. the irrelevant deformations in the IR to flow the geometry to a required UV geometry. Then in accordance with the chiral representation, the obtained numerical results of $\Psi_1,\Psi_2$ are utilised to determine the coefficients ${a_1}^i$ and ${a_2}^i$ in \eqref{UVbehaviour}. 
        
           Finally, {we should choose four sets of linearly independent horizon boundary conditions, i.e. the $a_+,a_-$ in \eqref{NearHorizon}, and then solve the equation of motion to get the coefficients ${a_1}^i$ and ${a_2}^i$ in \eqref{UVbehaviour}. Here we use $a_{1,2}^{i,I},a_{1,2}^{i,II},a_{1,2}^{i,III},a_{1,2}^{i,IV}$ to mark the coefficients corresponding to different horizon boundary conditions.}
           The expression of the Green's function can be defined with the following source and response matrix
           \be
           \label{CoefficientMatrix}
            M_s=\begin{pmatrix}
                {a_1}^{1,\mathrm{I}} & {a_1}^{1,\mathrm{II}} & {a_1}^{1,\mathrm{III}} & {a_1}^{1,\mathrm{IV}} \\
                {a_1}^{2,\mathrm{I}} & {a_1}^{2,\mathrm{II}} & {a_1}^{2,\mathrm{III}} & {a_1}^{2,\mathrm{IV}} \\
                {a_2}^{3,\mathrm{I}} & {a_2}^{3,\mathrm{II}} & {a_2}^{3,\mathrm{III}} & {a_2}^{3,\mathrm{IV}} \\
                {a_2}^{4,\mathrm{I}} & {a_2}^{4,\mathrm{II}} & {a_2}^{4,\mathrm{III}} & {a_2}^{4,\mathrm{IV}}
            \end{pmatrix},~M_r=\begin{pmatrix}
                -{a_2}^{1,\mathrm{I}} & -{a_2}^{1,\mathrm{II}} & -{a_2}^{1,\mathrm{III}} & -{a_2}^{1,\mathrm{IV}} \\
                -{a_2}^{2,\mathrm{I}} & -{a_2}^{2,\mathrm{II}} & -{a_2}^{2,\mathrm{III}} & -{a_2}^{2,\mathrm{IV}} \\
                {a_1}^{3,\mathrm{I}} & {a_1}^{3,\mathrm{II}} & {a_1}^{3,\mathrm{III}} & {a_1}^{3,\mathrm{IV}} \\
                {a_1}^{4,\mathrm{I}} & {a_1}^{4,\mathrm{II}} & {a_1}^{4,\mathrm{III}} & {a_1}^{4,\mathrm{IV}}
            \end{pmatrix}.
            \ee
            
            The Green's function can then be obtained as $G=i\Gamma^t M_r {M_s}^{-1}$, which can be utilised to {obtain the} topological Hamilton $H_ {eff}$. {Then we can use the effective Hamiltonian to calculate the topological invariant for the nodes that we have in the fermion spectral functions using the corresponding Berry curvature as in the weak coupling system.}

            {In summary, the steps for calculating the topological invariants in holographic system are:}
            \begin{enumerate}
            \item
            Given the background for a holographic model as \eqref{holoideal}, we use a pair of probe fermions to detect a topological invariant, whose action is e.g. \eqref{probeideal}. 
                \item Using given IR geometry \eqref{BackGroundIRGeometry} to impose the IR infalling boundary conditions for probe fermions \eqref{NearHorizon}.
                \item Solve the equations of motion \eqref{eom1} with the IR boundary conditions of the near horizon behaviour \eqref{NearHorizon} for the probe fermions. Then get the coefficients of the boundary asymptotic behaviour for probe fermions in \eqref{UVbehaviour}.
                \item After getting four linearly independent solutions, arrange the corresponding coefficients into the form \eqref{CoefficientMatrix}, and then we have the Green's function $G=i\Gamma^t M_r M_s^{-1}$.
                \item Define the topological Hamiltonian $H=-G^{-1}$, and then the corresponding Berry curvature can be used to get the topological invariant for the system as if the Hamiltonian is for a weak coupling system.
            \end{enumerate}
     
    \subsubsection{Fermion spectral functions and topological invariant for holographic Weyl semimetal}
    \label{wfs}
        In this subsection, to calculate the topological invariant for the holographic Weyl semimetal, we should give the spectrum function of probe fermions on the holographic Weyl semimetal to find out the location of Weyl nodes, so that we can integrate the Berry curvature around the nodes to get corresponding the Weyl charges.

        In the holographic model \eqref{holoideal}, the Weyl nodes are located along the axes $k_z$ with the ansatz \eqref{idealansatz}. 
        As previously stated, we can treat the topological Hamiltonian as an effective Hamiltonian as if we are working in a weakly coupled system, so the four branches of eigenvalues of the topological Hamiltonian can be viewed as four effective energy bands.
        From the standpoint of the energy band structure, the Weyl nodes are located at the specific value of $k_z$ where two bands cross together. In other words, two branches of eigenvalues for the topological Hamiltonian become zero simultaneously at this $k_z$.
        \begin{figure}
            \centering
            \includegraphics[width=0.815\textwidth]{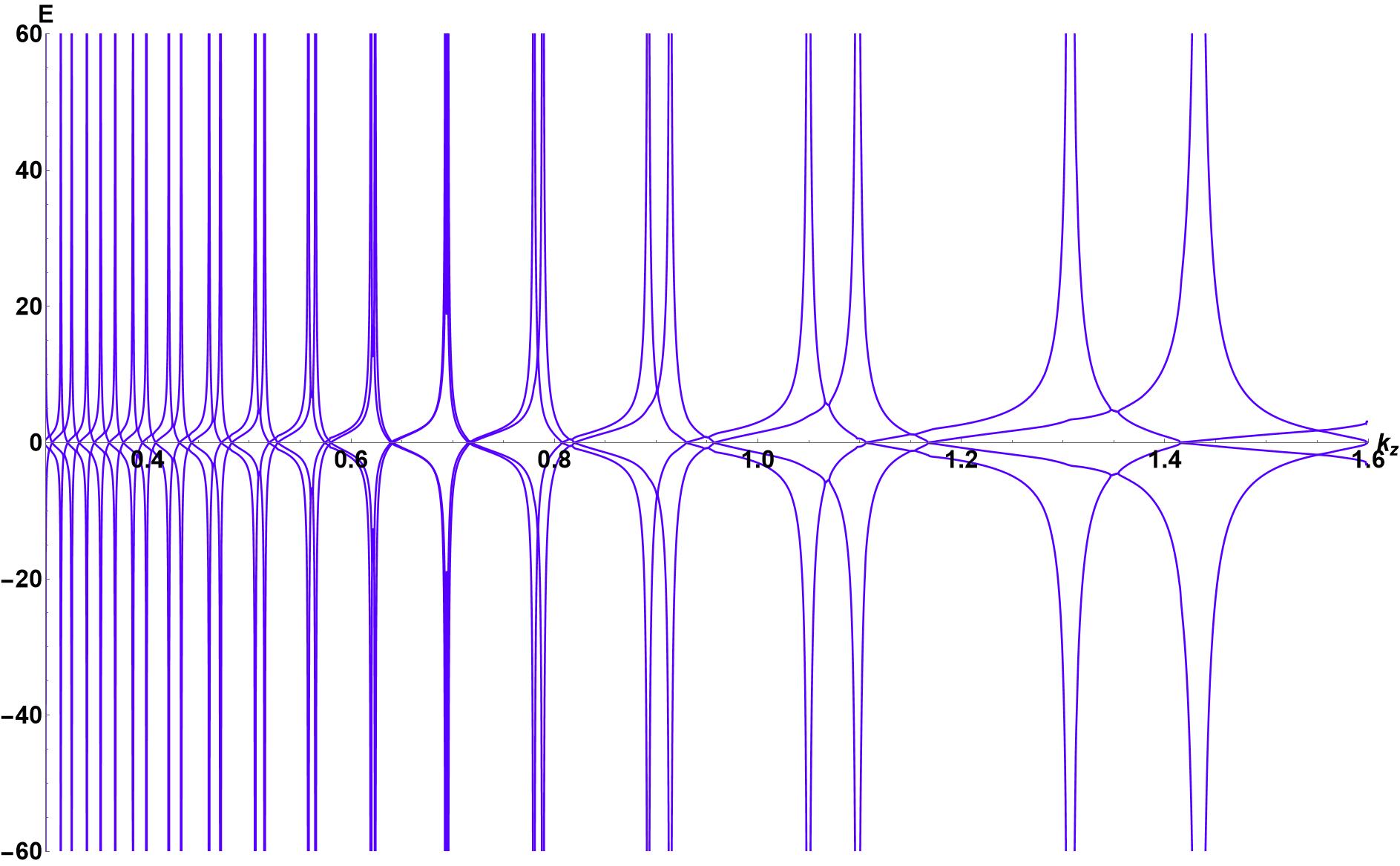}
            \caption{The four branches of eigenvalues for the topological Hamiltonian $H_{eff}(k)=-G^{-1}(0,k)$ along the $k_z$ axis for $\frac{M}{b}=0.705$. This can be viewed as the band structure of this effective Hamiltonian, therefore we use $E$ to label the vertical axis. The intersection points of these ``bands" with the $k_z$ axis are Weyl nodes. There are infinite many Weyl nodes whose distances between adjacent ones grow as $k_z$ increases for large $k_z$.}
            \label{fig2}
        \end{figure}
        
        
        After performing the required numerical calculations, we find that, different from the weak coupling case, there are now infinite many Weyl nodes along the $k_z$ axis, as shown in Fig. \ref{fig2}, where the energy dispersion, i.e. the eigenvalues of $H_{eff}$, is shown as a function of $k_z$\footnote{We set $k_x=k_y=0$ since the Weyl nodes are located solely in the $k_z$ direction.}. The existence of infinite many Weyl nodes is consistent with the multiple Fermi surfaces found in previous many holographic models, e.g. RN metal \cite{Liu:2009dm}, semi-local quantum liquid \cite{Iqbal:2011in}, electron star \cite{Hartnoll:2010gu, Hartnoll:2011dm, Cubrovic:2011xm}, BCS star\cite{Liu:2014mva}, and the holographic nodal line semimetal \cite{Liu:2018djq}. Mathematically, this multiple 
        Fermi surface behavior could be associated with the multiple bound states in the effective Schrödinger-like problem of solving Dirac fermions in the asymptotic AdS background. This existence of multiple Fermi surfaces is a special feature of holographic systems that cannot be adequately obtained by merely calculating the anomalous Hall conductivity or other transport parameters.

        With infinite many nodes, it is important to use the topological invariant to detect which are trivial ones and which are nontrivial ones. Trivial topological invariant for a Weyl node indicates an accidental band crossing which is not topologically protected, i.e. it could become a gap under a small perturbation. Nontrivial topological invariant, which is $\pm 1$ for a Weyl node, denotes a band crossing that is topologically protected and cannot be gapped under small perturbations. For a pair of Weyl nodes, their topological invariants are opposite with each other. Therefore, it is also important to indicate which two Weyl nodes are a pair of nodes that could be annihilated to form a Dirac node. This could be analyzed by studying the band structures for a set of values of $\frac{M}{b}$ up to the Dirac critical point, which be presented in a future work\cite{future}.

        Here we use the combined method of topological invariants and the observation of the band structure to distinguish Weyl nodes with opposite Weyl charges. The detailed analysis will be shown in the next section for the holographic $Z_2$ Weyl semimetal, which has the same structure as the Weyl nodes in the current holographic Weyl semimetal model. We summarize the result here. All the Weyl nodes in Figure. \ref{fig2} have nontrivial topological charges of $+1$ or $-1$, consistent with the fact that all band crossings at the $k_z$ axis are not accidental touchings. Regarding which two Weyl nodes could be viewed as a pair of nodes with opposite charges, we could show from the result of the topological invariants that the two nearest adjacent ones have opposite Weyl charges. However, there is a subtle point here. As illustrated in Fig. \ref{fig2}, upon increasing the momentum $k_z$ from zero, the distances between a specific pair of Weyl nodes initially decrease to (almost)zero and subsequently increase rapidly. This phenomenon can be substantiated through the calculation of the results of the topology invariant, a subject that will be addressed in Sec. \ref{sec:4}. As we could see from Figure. \ref{fig2}, this due to a band interchanging in the region between $k_z=0.6$ and $k_z=0.8$, and to the left of this region, the Weyl charges of the two nearest adjacent Weyl nodes display as $-1$,$+1$ while to the right of this region, the two charges are $+1$, $-1$ due to the band interchanging.

%% file: Section/Section3.tex
\section{Fermion spectral functions of the Weyl-$\boldmath Z_2$ semimetal}
\label{sec:3}
The electron is known to exhibit not only the degree of freedom of chirality but also the spin degree of freedom. The spin degree of freedom is also of significance to the topological properties of the many-body system, as evidenced by the occurrence of the spin Hall effect in graphene\cite{km1}. In consideration of the structural similarity of the energy bands of the Weyl semimetal to those of graphene, this many-body system is regarded as a natural candidate for the study of spin-related properties in three dimensions. However, the ideal Weyl semimetal primarily focuses on the chirality of the fermions. The Weyl-$\boldmath Z_2$ semimetal is a topological semimetal system that contains two pairs of nodes, such that both of the chirality and the spin degree of freedom can be taken into consideration\cite{Ji:2021aan}. The Weyl-$\boldmath Z_2$ semimetal is regarded as a rudimentary extension of the ideal Weyl semimetal, exhibiting divergent topological properties. Since the topology properties of a many-body system is determined by the energy band structures, this section will henceforth focus on an investigation of the fermion spectral functions for the Weyl-$\boldmath Z_2$ semimetal. The aim of this investigation is to demonstrate the similarities and differences in the energy band structures of the ideal and the Weyl-$\boldmath Z_2$ semimetals.

    \subsection{Review of the Weyl-$\boldmath Z_2$ semimetal}
\begin{figure}[h]
            \begin{minipage}{0.475\linewidth}
                \vspace{3pt}
                \centerline{\includegraphics[width=\textwidth]{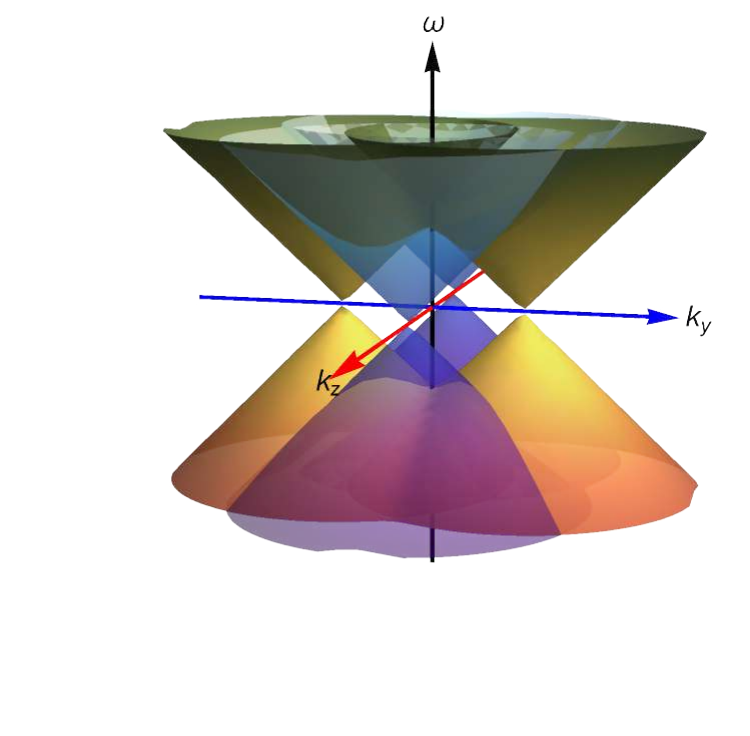}}
            \end{minipage}
            \begin{minipage}{0.5\linewidth}
                \vspace{3pt}
                \centerline{\includegraphics[width=\textwidth]{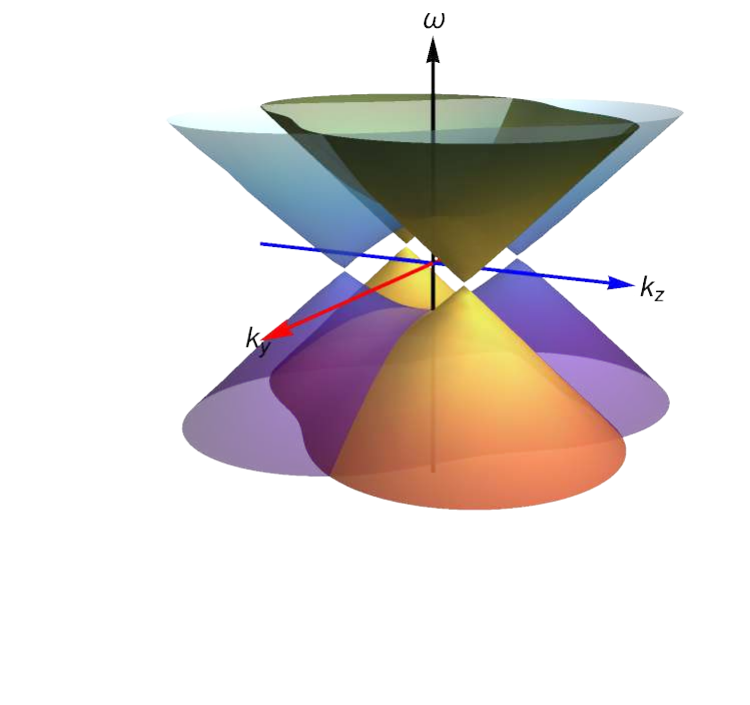}}
            \end{minipage}
            \caption{The band structure and the distribution of the nodes in the effective field theory of the Weyl-$\boldmath Z_2$ semimetal. The left panel is one type of Weyl-$\boldmath Z_2$ semimetal in which all the nodes are located on the $k_y$ and $k_z$ axis and the right panel is another type where the nodes locate on the quadrant bisector of $k_y-k_z$ plane.}
            \label{fig3}
        \end{figure}

    As previously stated, the Weyl-$\boldmath Z_2$ semimetal is characterised by the presence of two pairs of nodes with both Weyl and $Z_2$ charges. Consequently, the location of these nodes in the momentum space is of significance, as it has the capacity to exert an influence on the topological properties. Depending on this location of nodes, there could be two types of Weyl-$\boldmath Z_2$ semimetal, whose topological charges also have slight difference as shown in Fig. \ref{fig3}. In this work, we focus on the left type of Fig. \ref{fig3} where the two pairs of nodes are located in two directions perpendicular to each other in momentum space and the corresponding topological charges should be $(\pm 1,0)$ and $(0,\pm 1)$ for the Weyl and $\boldmath Z_2$ charges.

 \subsubsection{The effective field theory model of the Weyl-$\boldmath Z_2$ semimetal}
            In this part, an effective field theory model will be reviewed that has the capacity to describe the Weyl-$\boldmath Z_2$ semimetal shown in the left panel of Fig. \ref{fig3}. We have shown that through adding a Lorentz breaking term to the fundamental Quantum Electrodynamics (QED) process can describe an ideal Weyl semimetal, which is 
            \begin{equation}
                \mathcal{L}=\Bar{\psi}(i\gamma^\mu\partial_\mu-e\gamma^\mu A_\mu-b_\mu\gamma^\mu\gamma^5+m)\psi,
            \end{equation}
            where $\psi$ represents a four-component spinor, $A_\mu$ denotes the electromagnet gauge field and $b_\mu$ is associated with the axial gauge field. The Weyl-$\boldmath Z_2$ semimetal  in this framework \textcolor{purple}{can} be obtained by generalizing the Lagrangian of the ideal Weyl semimetal with the four-component spinor into a system of eight-component spinors as follows \cite{Ji:2021aan}
            \begin{equation}
            \label{WeylZ2SemimetalEffectiveLagrange}
            \mathcal{L}=\Psi^{\dagger}\left[\Gamma^{0}\left(i\Gamma^{\mu}\partial_{\mu}-e\Gamma^{\mu}A_{\mu}-\Gamma^{\mu} \Gamma^{5}b_{\mu} {\mI}_{1}+ {M}_{1} {\mI}_{1}+ {M}_{2} {\mI}_{2} \right)+\hat{\Gamma}^{0}\left(e\hat{\Gamma}^{\mu}\hat{A}_{\mu}- \hat{\Gamma}^{\mu}\hat{\Gamma}^{5}c_{\mu} {\mI}_{2}\right)\right]\Psi,
            \end{equation}
            where $\Psi$ is an eight-component spinor, and the new generalized `Gamma' matrices are defined as
            \be
            \Gamma^\mu:=\gamma^\mu\otimes\mathbb{I}_2,~~\Gamma^5:&=&\gamma^5\otimes\mathbb{I}_2,~~\hat{\Gamma}^\mu:=\gamma^\mu\otimes\mathbb{Z}_2,~~\hat{\Gamma}^5:=\gamma^5\otimes\mathbb{Z}_2,
            \ee
            with
            \be
            \mathbb{I}_2:&=&\begin{pmatrix}1 & 0 \\ 0 & 1 \end{pmatrix},~~~\mathbb{Z}_2:=\begin{pmatrix}1 & 0 \\ 0 & -1 \end{pmatrix},\nonumber\\
            \mathbf{I}_1:&=&diag(1,0,1,0,1,0,1,0),~~~\mathbf{I}_2:=diag(0,1,0,1,0,1,0,1).
            \ee
            $A_\mu$ is the electromagnet gauge field, $b_\mu$ is the chiral gauge field, while $\hat{A}_\mu$ is fictitious spin gauge field and $c_\mu$ is the so-called $\boldmath Z_2$ gauge field. The energy spectrum can be obtained by solving \eqref{WeylZ2SemimetalEffectiveLagrange}. Without loss of generality, we set $b_{\mu}=b\delta^{z}_{\mu}$ and $c_{\mu}=c\delta^{y}_{\mu}$, under which ansatz the Weyl nodes are located on the $z$ axis and the $\mathbb{Z}_2$ nodes are located on the $y$ axis.
            Here are the energy eigenvalues corresponding to the eight energy bands of system \eqref{WeylZ2SemimetalEffectiveLagrange} on the $k_x=0$ plane
            \be
          \label{z2spetrum} E_{1}=\pm\sqrt{\left(b_{z}\pm\sqrt{k_{z}^{2}+M_1^{2}}\right)^{2}+k_{y}^{2}},
            ~~~E_{2}=\pm\sqrt{\left(c_{y}\pm\sqrt{k_{y}^{2}+M_2^{2}}\right)^{2}+k_{z}^{2}}.
            \ee
            From \eqref{z2spetrum}, we deduce that two independent pairs of Weyl nodes exist within this system. The coordinates of these nodes are given by $(k_x, k_y, k_z)=\left(0,0,\pm\sqrt{b^{2}-M_1^{2}}\,\right)$ and $(k_x, k_y, k_z)=\left(0\,,\pm\sqrt{c^{2}-M_{2}^{2}},0\right).$
            Therefore, it can be concluded that, in the model \eqref{WeylZ2SemimetalEffectiveLagrange}, the two lines that link each independent pair of nodes are perpendicular to each other, as illustrated in the left panel of Fig. \ref{fig3}.

        \subsubsection{The holographic model of the Weyl-$\boldmath Z_2$ semimetal}
            In this section, a review of a holographic Weyl-$\boldmath Z_2$ semimetal model \cite{Ji:2021aan} is presented. This model is employed to describe scenarios where the quasiparticle picture is rendered invalid; that is to say, when the system {is strong coupled}. In accordance with the dictionary of the gauge gravity duality and in consideration of the holographic ideal Weyl semimetal \eqref{holoideal}, the action of a holographic Weyl-$\boldmath Z_2$ semimetal model can be expressed as follows
       \begin{align}
S&=\int d^5x \sqrt{-g}\bigg[\frac{1}{2\kappa^2}\Big(R+\frac{12}{L^2}\Big)+\frac{\alpha}{3}\epsilon^{abcde}A_a(3{F_V}_{bc}{F_V}_{de}+{F_A}_{bc}{F_A}_{de}+3{\hat{F_V}}_{bc}{\hat{F_V}}_{de}+{\hat{F_A}}_{bc}{\hat{F_A}}_{de})\nonumber\\
&+\frac{2\beta}{3}\epsilon^{abcde}\hat{A}_a(3\hat{F_V}_{bc}{F_V}_{de}+\hat{F_A}_{bc}{F_A}_{de})-\frac{1}{4}F^2-\frac{1}{4}\hat{F}^2-\frac{1}{4}F_5^2-\frac{1}{4}\hat{F}_5^2\nonumber\\
&-(D^{a}\Phi_{1})^{*}(D_{a}\Phi_{1})-(\hat{D}^{a}\Phi_{2})^{*}(\hat{D}_{a}\Phi_{2})-m|\Phi_1|^2-\frac{\lambda_1}{2}|\Phi_1|^4-m|\Phi_2|^2-\frac{\lambda_2}{2}|\Phi_2|^4\bigg]\,. \label{holoz2}
\end{align}                 
As a {generalization} from the holographic Weyl semimetal \eqref{holoideal}, the model \eqref{holoz2} introduces two additional gauge fields: $\hat{V}$ with the field strength $\hat{F}_V$ and $\hat{A}$ with the field strength $\hat{F}_A$, respectively. This is due to the fact that the spin gauge field and the $\boldmath Z_2$ gauge field are necessary elements in a Weyl-$\boldmath Z_2$ semimetal. Two Chern-Simons terms, with coupling constants $\alpha$ and $\beta$, respectively, have been shown to produce the chiral anomaly and the $\boldmath Z_2$ axial anomaly.  Scalar field $\Phi_1$ (or $\Phi_2$) is charged only under the axial (or $\boldmath Z_2$ axial) gauge symmetries with the corresponding covariant derivative as $D_a:=\partial_a-iq_1A_a$(or $\hat{D}_a:=\partial_a-iq_2\hat{A}_a$).  We have also set the scalar bulk mass to be $m^2L^2=-3$ as in \eqref{holoideal}.

            \eqref{holoz2} can be solved at zero temperature with the ansatz
            \be
            \label{z2ansatz}
            ds^2&=&u(r)(-dt^2+dx^2)+f(r) dy^2+h(r) dz^2+\frac{dr^2}{u(r)},~\Phi_1=\phi_1(r),~\Phi_2=\phi_2(r)\nonumber\\ A&=&A_z(r)dz,~V=0,~\hat{A}=C_y(r)dy,~\hat{V}=0,
            \ee
            where $u,f,h,\phi_1,\phi_2,A_z,C_y$ are all real functions of $r$. As $r\to \infty$, that is to say, the AdS boundary, the {corresponding asymptotic behaviour is the following}
            \begin{equation}
                u\to r^2,~f\to r^2,~h\to r^2,~\phi_1\to\frac{M_1}{r},~\phi_2\to\frac{M_2}{r},~A_z\to b,~C_y\to c.
            \end{equation}
           At $r \to 0$, i.e. the horizon, the topologically nontrivial Weyl-$\boldmath Z_2$ semimetal phase has the following IR geometry  
            \be
            \label{WeylZ2BackgroundGeometry}
              u&=&f=h=r^2\nonumber\\
                \phi_1&=&\phi_{10}\sqrt{\frac{\pi}{8}}\left(\frac{q_1a_0}{r}\right)^{\frac{3}{2}}e^{-\frac{q_1 a_0}{r}},~\phi_2=\phi_{20}\sqrt{\frac{\pi}{8}}\left(\frac{q_2c_0}{r}\right)^{\frac{3}{2}}e^{-\frac{q_2 c_0}{r}}\nonumber\\
                A_z&=&a_0+\frac{\pi a_0^2 \phi_{10}^2}{16 r}e^{-\frac{2 q_1 a_0}{r}},~C_y=c_0+\frac{\pi c_0^2 \phi_{20}^2}{16 r}e^{-\frac{2 q_2 c_0}{r}},
            \ee
            where $a_0,c_0,\phi_{10},\phi_{20}$ are shooting parameters. Once the appropriate values for the shooting parameters have been selected, {we can use the numerical method to solve the equation of motion of $u,f,h,\phi_1,\phi_2,A_z,C_y$ with the initial condition \eqref{WeylZ2BackgroundGeometry}}.

    \subsection{Probe fermions on the holographic Weyl-$\boldmath Z_2$ semimetal}
        As previously stated, the energy band structure of the holographic topological system can be obtained by probing the system with fermions. In the case of the holographic Weyl semimetal, the employment of two probe fermions is crucial to describe opposite chiralities. Conversely, in a Weyl-$\boldmath Z_2$ semimetal, each node possesses not only a chiral charge but also a $\boldmath Z_2$ charge. Consequently, two pairs of probe fermions are required in the Weyl-$\boldmath Z_2$ semimetal. One pair ($\Psi_1,\Psi_2$) is employed to correspond to boundary fermions with opposite chiralities, with the topological invariant being the Weyl charge (i.e. the chiral charge). The other pair {($\Psi_3,\Psi_4$)} corresponds to the boundary spin analog of opposite chiral charges, associated with the $\boldmath Z_2$ symmetry with the topological invariant being the $\boldmath Z_2$ charge.
        
        Based on this, the probe fermion's action for the Weyl-$\boldmath Z_2$ semimetal has the form:
        \be
        \label{z2probe1}
        S=S_1+S_2+S_3+S_4+S_\Phi,
        \ee
        where\footnote{In the  Weyl-$\boldmath Z_2$ semimetal, the components $\Psi_{1,2}$ are defined as eight-component spinors. Consequently, the corresponding `Gamma' matrix must also be redefined as an $8\times8$ matrix, given by: $\Gamma^a:=\Gamma^a\otimes\mathbb{I}_2,~\hat{\Gamma}^a:=\Gamma^a\otimes\mathbb{Z}_2$. For the sake of unified expression with the Weyl semimetal case, the introduction of new symbols for the `Gamma' matrix is not warranted here.}
        \be
        \label{z2femionequation}
        S_1&=\int\sqrt{-g}d^5x\Bar{\Psi}_1[\Gamma^a( D_a-i A_a)-m_f]\Psi_1,\nonumber\\
        S_2&=\int\sqrt{-g}d^5x\Bar{\Psi}_2[\Gamma^a( D_a+iA_a)+m_f]\Psi_2\nonumber\\
        S_3&=\int\sqrt{-g}d^5x\Bar{\Psi}_3[\hat{\Gamma}^a( D_a-iC_a)-m_f]\Psi_3,\nonumber\\
        S_4&=\int\sqrt{-g}d^5x\Bar{\Psi}_4[\hat{\Gamma}^a( D_a+iC_a)+m_f]\Psi_4,
        \ee
        and
        \be
        S_\Phi=-\int\sqrt{-g}d^5x(\eta_1\Phi_1\Bar{\Psi}_1\Psi_2+\eta_1^*\Phi_1^*\Bar{\Psi}_ 2\Psi_1+\eta_2\Phi_2\Bar{\Psi}_ 3\Psi_4+\eta_2^*\Phi_2^*\Bar{\Psi}_4\Psi_3).
        \ee

        Similarly, the corresponding equation of motion or the probe fermions can be obtained by doing variation of \eqref{z2femionequation}. With the background geometry \eqref{z2ansatz} and the following ansatz
        \be
        \Psi=\psi(r)e^{ik_\mu x^\mu},~\Phi_1(r)=\phi_1,~\Phi_2(r)=\phi_2,~A_z(r)=A_zdz,~C_y(r)=C_y dy,
        \ee
        the equation of motions for the four probe fermions are
        \be
        &\sqrt{u}\Gamma^{{r}}\partial_r\psi_1+\left(-i\frac{\omega}{\sqrt{u}}\Gamma^{{t}}+i\frac{k_x}{\sqrt{u}}\Gamma^{{x}}+i\frac{k_y}{\sqrt{f}}\Gamma^{{y}}+i\frac{(k_z-q_1A_z)}{\sqrt{h}}\Gamma^{{z}}\right)\psi_1-m_f\psi_1-\eta_1\phi_1\psi_2=0,~~~~~~~\nonumber\\
        &\sqrt{u}\Gamma^{{r}}\partial_r\psi_2+\left(-i\frac{\omega}{\sqrt{u}}\Gamma^{{t}}+i\frac{k_x}{\sqrt{u}}\Gamma^{{x}}+i\frac{k_y}{\sqrt{f}}\Gamma^{{y}}+i\frac{(k_z+q_1A_z)}{\sqrt{h}}\Gamma^{{z}}\right)\psi_2+m_f\psi_2-\eta_1\phi_1\psi_1=0,~~~~~~~\nonumber\\
        &\sqrt{u}\hat{\Gamma}^{{r}}\partial_r\psi_3+\left(-i\frac{\omega}{\sqrt{u}}\hat{\Gamma}^{{t}}+i\frac{k_x}{\sqrt{u}}\hat{\Gamma}^{{x}}+i\frac{(k_y-q_2C_y)}{\sqrt{f}}\hat{\Gamma}^{{y}}+i\frac{k_z}{\sqrt{h}}\hat{\Gamma}^{{z}}\right)\psi_3-m_f\psi_3-\eta_2\phi_2\psi_4=0,~~~~~~~\nonumber\\
        &\sqrt{u}\hat{\Gamma}^{{r}}\partial_r\psi_4+\left(-i\frac{\omega}{\sqrt{u}}\hat{\Gamma}^{{t}}+i\frac{k_x}{\sqrt{u}}\hat{\Gamma}^{{x}}+i\frac{(k_y+q_2C_y)}{\sqrt{f}}\hat{\Gamma}^{{y}}+i\frac{k_z}{\sqrt{h}}\hat{\Gamma}^{{z}}\right)\psi_4+m_f\psi_4-\eta_2\phi_2\psi_3=0.~~~~~~~
        \ee
        At $r\to 0$ in the vicinity of the horizon, the background fields have the following behaviour
        \be
        \label{irbeh}
        u=r^2,~~ f=r^2,~~ h=r^2,~~\phi_1=0,~~\phi_2=0,~~A_z=a_0,~~C_y=c_0.
        \ee
        
        The solutions to the probe fermions in the vicinity of the horizon are a form analogous to that in Weyl semimetal:
        \be
        \label{z2probe}
        &{\psi_{1}}_{horizon}=\frac{1}{\sqrt{r}}K_{m+\frac{1}{2}}\left(\frac{k_1}{r}\right)a_+ +\frac{ik_{1\mu}\Gamma^\mu}{k_1}\frac{1}{\sqrt{r}}K_{m-\frac{1}{2}}\left(\frac{k_1}{r}\right)a_++\cdots,~a_+\in \Im\Gamma_+,\nonumber\\
        &{\psi_{2}}_{horizon}=\frac{1}{\sqrt{r}}K_{m+\frac{1}{2}}\left(\frac{k_2}{r}\right)a_- +\frac{ik_{2\mu}\Gamma^\mu}{k_2}\frac{1}{\sqrt{r}}K_{m-\frac{1}{2}}\left(\frac{k_2}{r}\right)a_-+\cdots,~a_-\in \Im\Gamma_-,\nonumber\\
        &{\psi_{3}}_{horizon}=\frac{1}{\sqrt{r}}K_{m+\frac{1}{2}}\left(\frac{k_3}{r}\right)b_+ +\frac{ik_{3\mu}\hat{\Gamma}^\mu}{k_3}\frac{1}{\sqrt{r}}K_{m-\frac{1}{2}}\left(\frac{k_3}{r}\right)b_++\cdots,~b_+\in \Im\hat{\Gamma}_+,\nonumber\\
        &{\psi_{4}}_{horizon}=\frac{1}{\sqrt{r}}K_{m+\frac{1}{2}}\left(\frac{k_4}{r}\right)b_- +\frac{ik_{4\mu}\hat{\Gamma}^\mu}{k_4}\frac{1}{\sqrt{r}}K_{m-\frac{1}{2}}\left(\frac{k_4}{r}\right)b_-+\cdots,~b_-\in \Im\hat{\Gamma}_-,
        \ee
        where $\Gamma_\pm:=\frac{1\pm \Gamma^r}{2},~\hat{\Gamma}_\pm:=\frac{1\pm\hat{\Gamma}^r}{2}$, {$\cdots$ denotes higher order terms}, and $k_1=(-\omega,k_x,k_y,k_z-q_1 a_0),~k_2=(-\omega,k_x,k_y,k_z+q_1 a_0),~k_3=(-\omega,k_x,k_y-q_2 c_0,k_z),~k_4=(-\omega,k_x,k_y+q_2 c_0,k_z)$. It is possible to derive the initial condition of the coupled first-order Dirac equations \eqref{z2femionequation} using equation \eqref{z2probe}, as was previously demonstrated in Sec. \ref{sec:2}.
        
        In the near-UV domain, the spinors can be divided into two parts: one of which is treated as the source, with components proportional to $r^{m_f}$, and the other as the response, with components proportional to  $r^{-m_f}$. The spinors can thus be expressed in the following manner when the Gamma matrices are defined in chiral representation
        \be
        \label{z2uvc}
        &\psi_1\stackrel{r\to\infty}{\longrightarrow}\begin{pmatrix}{a_1}^1r^{m_f}\\ {a_1}^2r^{m_f}\\ {a_1}^3r^{m_f}\\ {a_1}^4r^{m_f}\\ {a_1}^5r^{-m_f}\\ {a_1}^6r^{-m_f}\\ {a_1}^7r^{-m_f}\\ {a_1}^8r^{-m_f}\end{pmatrix},~~~~~\psi_2\stackrel{r\to\infty}{\longrightarrow}\begin{pmatrix}{a_2}^1r^{-m_f}\\ {a_2}^2r^{-m_f}\\ {a_2}^3r^{-m_f}\\ {a_2}^4r^{-m_f}\\ {a_2}^5r^{m_f}\\ {a_2}^6r^{m_f}\\ {a_2}^7r^{m_f}\\ {a_2}^8r^{m_f}\end{pmatrix},
        \ee
        \be
        \label{z2uvz}
        &\psi_3\stackrel{r\to\infty}{\longrightarrow}\begin{pmatrix}{a_3}^1r^{m_f}\\ {a_3}^2r^{-m_f}\\ {a_3}^3r^{m_f}\\ {a_3}^4r^{-m_f}\\ {a_3}^5r^{-m_f}\\ {a_3}^6r^{m_f}\\ {a_3}^7r^{-m_f}\\ {a_3}^8r^{m_f}\end{pmatrix},~~~~~\psi_4\stackrel{r\to\infty}{\longrightarrow}\begin{pmatrix}{a_4}^1r^{-m_f}\\ {a_4}^2r^{m_f}\\ {a_4}^3r^{-m_f}\\ {a_4}^4r^{m_f}\\ {a_4}^5r^{m_f}\\ {a_4}^6r^{-m_f}\\ {a_4}^7r^{m_f}\\ {a_4}^8r^{-m_f}\end{pmatrix}.
        \ee        
        
        The four spinors can be categorised into two distinct groups. The combination of the $\psi_1$ and $\psi_2$ spinors can be employed to detect the Weyl charge. The Green's function corresponding to the chirality of the system can be written as $G_{chirality}=\underset{r\to\infty}{\lim} r^{2m_f}\Gamma^t\Xi_{chirality}$, where $\Xi_{chirality}$ is calculated through $\chi_{chirality}=-i\Xi_{chirality}\Psi_{chirality}$. $\Psi_{chirality}$ and $\chi_{chirality}$ are defined as the combination of $\psi_1$ and $\psi_2$, which are
        \be
        \Psi_{chirality}=\Gamma_{+}\psi_1+\Gamma_{-}\psi_2,~~~~\chi_{chirality}:=\Gamma_{-}\psi_1-\Gamma_+\psi_2.
        \ee

        In a parallel manner, the combination of $\psi_3$ and $\psi_4$ are utilised to examine the possible $\boldmath Z_2$ charge (also known as the `spin charge') of the system. The corresponding Green's function can be expressed as $G_{\boldmath Z_2}=\underset{r\to\infty}{\lim} r^{2m_f}\hat{\Gamma}^t\Xi_{\boldmath Z_2}$, where $\Xi_{\boldmath Z_2}$ is derived through the calculation of $\chi_{\boldmath Z_2}=-i\Xi_{\boldmath Z_2}\Psi_{\boldmath Z_2}$. $\Psi_{\boldmath Z_2}$and $\chi_{\boldmath Z_2}$ are defined as
        \be
        \Psi_{\boldmath Z_2}=\hat{\Gamma}_{+}\psi_3+\hat{\Gamma}_{-}\psi_4,~~~~\chi_{\boldmath Z_2}:=\hat{\Gamma}_{-}\psi_3-\hat{\Gamma}_+\psi_4.
        \ee

        The establishment of the Green's function, $G_{chirality}$ and $G_{\boldmath Z_2}$, enables the subsequent calculation of the topological Hamiltonian through the implementation of the frequency, $\omega$, set to zero. In the following, the energy spectrum of the topological Hamiltonian will be presented first, and the calculation of the topological invariant, i.e. Weyl charge and $\boldmath Z_2$ charge, will be left to Sec. \ref{sec:4}.
        
        The topological Hamiltonian is known to comprise all the topological information of the system. In the case of the Weyl-$\boldmath Z_2$ semimetal, which has both chiral and $\boldmath Z_2$ degrees of freedom, it is expected that two distinct effective Hamiltonians would be better at manifesting the topology properties of the system than a single one. With this consideration, two effective Hamiltonians are defined as $H_{Weyl}=-{G_{chirality}}^{-1}(0,\mathbf{k})$ and ${H}_{\boldmath Z_2}=-{G_{\boldmath Z_2}}^{-1}(0,\mathbf{k})$, respectively. The respective energy spectrum (the effective band structure) and topological invariant can then be obtained by calculating the eigenvalues of them.

        From a different perspective, the holographic model \eqref{holoz2} encompasses two unconstrained dimensionless parameters, namely $\frac{M_1}{b}$ and $\frac{M_2}{c}$. The phase diagram can be specified by these two parameters\footnote{Another parameter $\frac{c}{b}$ has been shown not to exert any influence on the topology properties of the system. Therefore, it can be fixed to be one without loss of generality.}. 
        With the other parameters in the model fixed as $q_1=q_2=1,~2\kappa^2=1,~\alpha=\beta=1,~\lambda_1=\lambda_2=\frac{1}{10}$, the phase where two pairs of nodes exist is confined to  the region $\frac{M_1}{b}<0.908$ and $\frac{M_2}{c}<0.908$ \cite{Ji:2021aan}.

        It is evident that there is a strong similarity between the chiral and the $\boldmath Z_2$ degrees of freedom. We could verify this point in the limit $\frac{M_1}{b}\to 0, \frac{M_2}{c}\to 0$ by investigating the energy spectrum of the topological Hamiltonian in this limit. The corresponding Green's functions can be obtained by employing the near horizon solution \eqref{z2probe}, which is
        \be
        \label{MOverBToZeroLimitGreensFunction}
        G_{chirality}&=-\Gamma^t\left(\frac{k_1^{2m-1}}{4^m}\frac{\Gamma\left(\frac{1}{2}-m\right)}{\Gamma\left(\frac{1}{2}+m\right)}k_{1\mu}\Gamma^{\mu}\Gamma_{+}-\frac{k_2^{2m-1}}{4^m}\frac{\Gamma\left(\frac{1}{2}-m\right)}{\Gamma\left(\frac{1}{2}+m\right)}k_{2\mu}\Gamma^{\mu}\Gamma_{-}\right),\nonumber\\
        G_{\boldmath Z_2}&=-\hat{\Gamma}^t\left(\frac{k_3^{2m-1}}{4^m}\frac{\Gamma\left(\frac{1}{2}-m\right)}{\Gamma\left(\frac{1}{2}+m\right)}k_{3\mu}\hat{\Gamma}^{\mu}\hat{\Gamma}_{+}-\frac{k_4^{2m-1}}{4^m}\frac{\Gamma\left(\frac{1}{2}-m\right)}{\Gamma\left(\frac{1}{2}+m\right)}k_{4\mu}\hat{\Gamma}^{\mu}\hat{\Gamma}_{-}\right).
        \ee
        \begin{figure}[h]
        \begin{minipage}{0.5\linewidth}
                \vspace{3pt}
                \centerline{\includegraphics[width=0.8\textwidth]{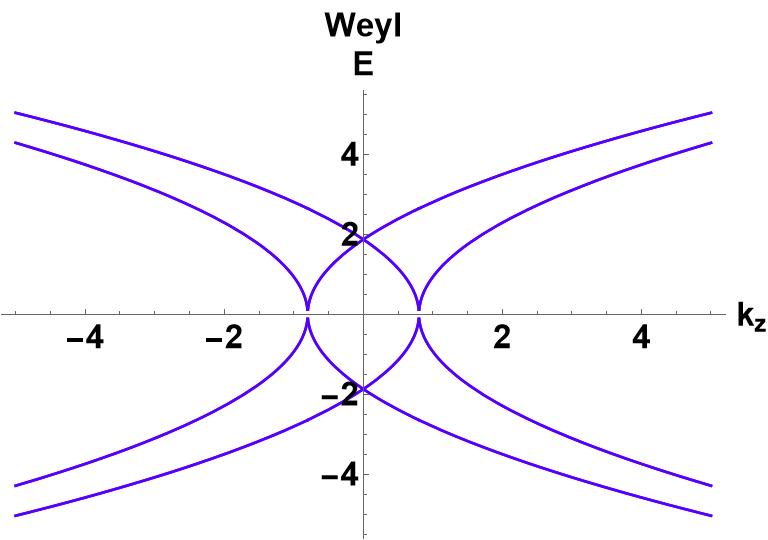}}
            \end{minipage}  \begin{minipage}{0.5\linewidth}
                \vspace{3pt}
                \centerline{\includegraphics[width=0.8\textwidth]{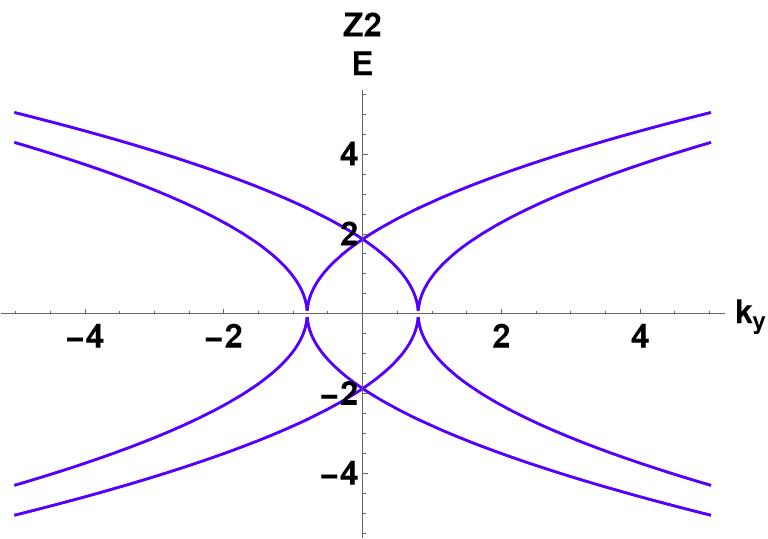}}
            \end{minipage}
            \caption{The spectral function of topological Hamiltonian $H_{Weyl}=-{G_{chirality}}^{-1}${, which has been represented in \eqref{MOverBToZeroLimitGreensFunction}(left panel) along $k_z$ axis} and $H_{\boldmath Z_2}=-{G_{\boldmath Z_2}}^{-1}$(right panel) {along $k_y$ axis} at the limit $\frac{M_1}{b}\to 0, \frac{M_2}{c}\to 0$ with the parameter $b=c=0.8$ for the holographic Weyl $\boldmath{Z}_2$ semimetal. That is, the eigenvalues for corresponding topological Hamiltonian as a function of $k_z$(or $k_y$). Note that the distances between two nodes are determined by the parameter $b$ and $c$.}
            \label{fig4}
        \end{figure} 
        
        From the Green's functions we can obtain the topological Hamiltonian and the band structure accordingly.
        In consideration of the ansatz \eqref{z2ansatz}, it can be posited that a pair of Weyl nodes will be situated on the $k_z$ axis which carry the non-zero chiral charge, and a pair of  $\boldmath Z_2$ nodes will be located on the $k_y$ axis which carry the non-zero $\boldmath Z_2$ charge. Therefore, it would be reasonable to figure out the band structure of $H_{Weyl}$ along the $k_z$ axis and $H_{\boldmath Z_2}$ along the $k_y$ axis.
        As illustrated in Fig. \ref{fig4}, the left panel corresponds to the energy spectrum of the $H_{Weyl}$ along the $k_z$ axis, while the right panel represents the spectrum of the $H_{\boldmath Z_2}$ along the $k_y$ axis in the limit $\frac{M_1}{b}\to 0, \frac{M_2}{c}\to 0$. The spectrum verifies the presence of two pairs of nodes, located independently along the axes perpendicular to each other. 
        In this limit, the distances between the pairs of nodes are determined by the values of $b$ and $c$. This also verifies the similarity between the band structures of the chiral and $\boldmath Z_2$ degrees of freedom from $H_{Weyl}$ and $H_{\boldmath Z_2}$.
       
\begin{figure}[h!]
            \includegraphics[width=\textwidth]{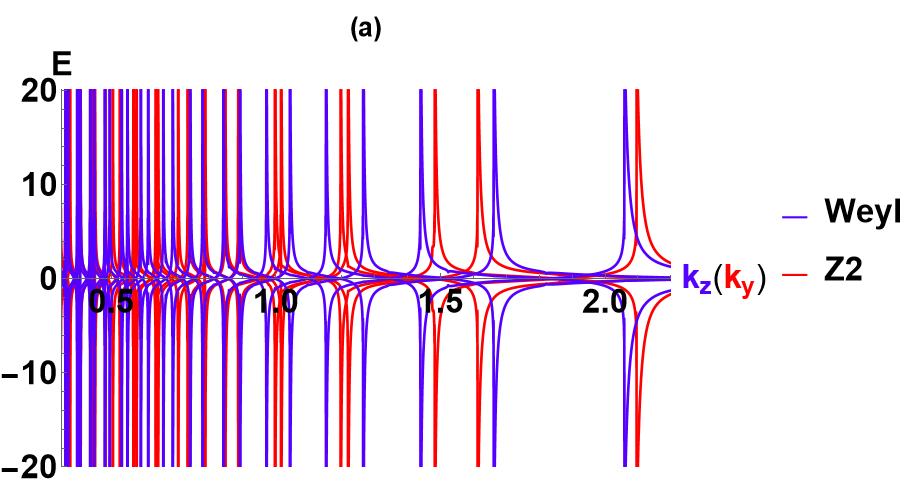}
            \begin{minipage}{0.5\textwidth}
                \vspace{3pt}
                \centerline{\includegraphics[width=\textwidth]{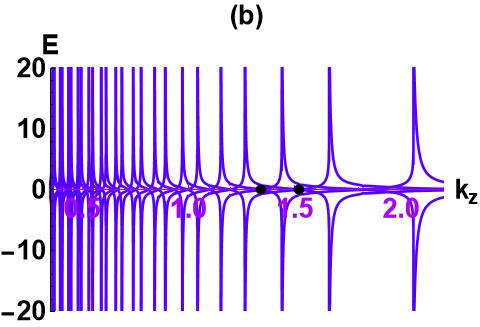}}
            \end{minipage}  
            \begin{minipage}{0.5\textwidth}
                \vspace{3pt}
                \centerline{\includegraphics[width=\textwidth]{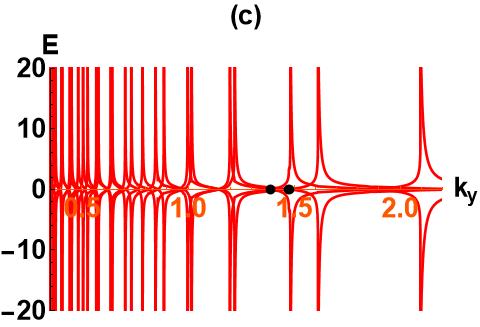}}
            \end{minipage}
            
            \caption{Eight branches of eigenvalues for topological Hamiltonian $H_{Weyl}$(blue line) and $H_{\boldmath Z_2}$(red line) at the value $\frac{M_1}{b}=0.0463, \frac{M_2}{c}=0.0494$. This can be viewed as the band structure of this effective Hamiltonian, therefore we use $E$ to label the vertical axis. Note that all the bands are doubly degenerate. The two nodes denoted by black dots in (b) can be considered as a pair and the two nodes denoted by black dots in (c) can be considered as a pair which could annihilate into a Dirac point at the critical  topological phase transition point.} 
            \label{fig5}
        \end{figure}

        In the following discussion we will consider the case $\frac{M_1}{b},\frac{M_2}{c}\ne 0$ in which backreaction have to be taken into consideration, so numerical methods would be employed. Since we have pointed out that when $\frac{M_1}{b}=\frac{M_2}{c}$ the band structures for $H_{Weyl}$ and $H_{\boldmath Z_2}$ will be the same, we will deal with the situation $\frac{M_1}{b}\ne\frac{M_2}{c}$. Without loss of generality, the value of $\frac{M_1}{b}$ is set to be 0.0463 and $\frac{M_2}{c}=0.0494$. Then the corresponding spectrums are presented in Fig. \ref{fig5}.
        
        As demonstrated in Fig. \ref{fig5}, the emergence of multiple nodes is also observed. The distances between a particular pair of nodes exhibit a similar trend when the momentum increases as those found in Fig. \ref{fig2} for the holographic Weyl semimetal.
        As there are multiple nodes now, it is important to determine which two nodes should form a pair of nodes with opposite topological charges. A way to examine this is to compare the distances between the nodes as $\frac{M}{b}$ changes and this will be confirmed by the calculation of topological invariant in the next section. When the value of $\frac{M}{b}$ increases up to the critical value $\left(\frac{M}{b}\right)_c$, the distance between the two nodes which form a pair will shrink to $0$, that is to say, two nodes become a Dirac node at the critical value $\left(\frac{M}{b}\right)_c$. Therefore as $\frac{M}{b}$ increases, we can identify with the adjacent two nodes between which the distance decreases as a pair.
        
            We can compare the two figures (b) and (c) in Fig. \ref{fig5}. As illustrated in model \eqref{holoz2}, at the critical phase two pairs of nodes merge to form a single Dirac node, the corresponding critical value is $\left(\frac{M_1}{b}\right)_c=\left(\frac{M_2}{c}\right)_c=0.908$\cite{Ji:2021aan}. Therefore, the distance between a specific pair of nodes will decrease as the values of $\frac{M_1}{b}$ or $\frac{M_2}{c}$ increase. Since $\frac{M_2}{c}>\frac{M_1}{b}$, (c) has a larger $\frac{M_2}{c}$, therefore it has a smaller distance between each pair of nodes compared to (b). From this point, we could see that the two adjacent nodes that we denote with two black dots in (b) or (c) form a pair of nodes and this will be confirmed by the calculation of topological invariants of the next section.

%% file: Section/Section4.tex
\section{Topological invariant of the  Weyl-$\boldmath{Z}_2$ semimetal}
\label{sec:4}
In contrast to the case of Weyl nodes, for which the Chern number naturally serves as the topological invariant, the $\boldmath{Z}_2$ degree of freedom has no naturally suitable candidate for a topological invariant. Nevertheless, the fact that the $\boldmath{Z}_2$ anomaly is employed to manifest the spin imbalance of two fermions, and the high similarity of the $\boldmath{Z}_2$ and the chiral degree of freedom, should be considered.
The spin-Chern number can then serve as the topological invariant for measuring the $\boldmath{Z}_2$ charge, and has been applied to several condensed matter systems. In this section, the spin-Chern number is reviewed first. Subsequently, the topological invariant of the holographic Weyl-$\boldmath{Z}_2$ semimetal with the effective topological Hamiltonians $H_{Weyl}$ and ${H}_{\boldmath Z_2}$ will be calculated.   
    \subsection{The spin-Chern number in the condensed matter systems}
        We first illustrate the definition of spin-$\boldmath Z_2$ charge in a typical condensed matter system, a hexagonal model. Note that in such a system with spin-orbit coupling, the whole system has a trivial Chern number or Weyl charge, but its subsystem has non-zero Chern number which can be utilized to define pseudo-spin Chern number. In order to point this fact explicitly, we start from following Hamiltonian\cite{Ezawa_2014}
            \be H=\hbar v_F(k_x\sigma^3\otimes I_2\otimes\sigma^1+k_y I_2\otimes I_2\otimes\sigma^2)+\lambda_{SO} \sigma^3\otimes\sigma^3\otimes\sigma^3+m_z I_2\otimes\sigma^3\otimes\sigma^3,\ee
        where $v_F$ denotes the Fermi velocity, $\lambda_{SO}$ denotes the spin-orbit coupling constant, and $m_z$ is the mass term which breaks time-reversal symmetry. Note that this Hamiltonian can commute with $\sigma^3\otimes I_2\otimes  I_2,~ I_2\otimes\sigma^3 \otimes I_2,$ and $\sigma^3\otimes \sigma^3\otimes I_2$. Without loss of generality, we can choose the common eigenvectors of $\sigma^3\otimes I_2\otimes  I_2$ and $I_2\otimes  \sigma^3\otimes I_2$ as a basis, so that the Hamiltonian can be written as four diagonal blocks, each of which can be denoted using the eigenvalues of $\sigma^3\otimes I_2\otimes  I_2$ and $I_2\otimes  \sigma^3\otimes I_2$ uniquely,
            \be H={\bf diag}\left(\begin{pmatrix}\lambda_{SO}+m_z & \hbar v_F(k_x-ik_y) \\ \hbar v_F(k_x+i k_y) & -\lambda_{SO}-m_z\end{pmatrix},~ \begin{pmatrix}-\lambda_{SO}-m_z & \hbar v_F(k_x-ik_y) \\ \hbar v_F(k_x+ik_y) & \lambda_{SO}+m_z\end{pmatrix},\right.\nonumber\\
            \left.\begin{pmatrix}-\lambda_{SO}+m_z & \hbar v_F(-k_x-ik_y) \\ \hbar v_F(-k_x+ik_y) & \lambda_{SO}-m_z\end{pmatrix},~\begin{pmatrix}\lambda_{SO}-m_z & \hbar v_F(-k_x-ik_y) \\ \hbar v_F(-k_x+ik_y) & -\lambda_{SO}+m_z\end{pmatrix}\right). \ee
        
            Each block has their own Chern number. Take the first block as an example,
            it corresponds to the eigenvalue $1$ for eigenvector $\sigma^3\otimes I_2\otimes  I_2$ and $1$ for eigenvector $I_2\otimes  \sigma^3\otimes I_2$
            , such that we can calculate the Chern number $C_{1,1}$ for the first block. The rest blocks have the Chern numbers $C_{1,-1},C_{-1,1}$ and $C_{-1,-1}$, respectively. Then a series of pseudo-spin Chern number can be discussed, such as total Chern numbers $C=C_{1,1}+C_{1,-1}+C_{-1,1}+C_{-1,-1}$, spin Chern number $C_{spin}=\frac{1}{2}(C_{1,1}-C_{1,-1}+C_{-1,1}-C_{-1,-1})$ and valley Chern number $C_{valley}=\frac{1}{2}(C_{1,1}+C_{1,-1}-C_{-1,1}-C_{-1,-1})$\cite{Ezawa_2014}.
        
            After discussing the example in hexagonal lattice with spin-orbit coupling, it is nature to give the definition for spin Chern number. Given a Hamiltonian $\hat{H}(k)$, if there is a spin operator $\sigma^z$ whose eigenvalue is $\pm 1$ and it can commute with the Hamiltonian $\hat{H}(k)$, the Hamiltonian can be transformed into a block diagonal matrix under spin operator's eigenvectors:
            \be\hat{H}(k)=\begin{pmatrix}\hat{H}_{\uparrow}(k) & 0\\ 0 & \hat{H}_{\downarrow}(k)\end{pmatrix},~\sigma^z=\begin{pmatrix}\mathbb{I} & 0 \\ 0 & -\mathbb{I}\end{pmatrix}.\ee
            
            $\hat{H}_{\uparrow}$ means that this block can be denoted with the eigenvalue $+1$ for spin operator $\sigma^z$, i.e. it corresponds to the spin up part of the system, and $\hat{H}_{\downarrow}$ corresponds to the spin down part respectively. Each block $\hat{H}_{\uparrow},\hat{H}_{\downarrow}$ has their own Weyl charge $C_\uparrow,C_\downarrow$ \footnote{In the hexagonal system, one can consider the operator $\mathbb{I}_2\otimes\sigma^3\otimes\mathbb{I}_2$ as the spin operator, so that $C_{\uparrow}=C_{1,1}+C_{-1,1}$ and $C_{\downarrow}=C_{1,-1}+C_{-1,-1}$.} In order to shed light on the difference between the spin up and spin down part of the system, the $\boldmath Z_2$ topological invariant can be defined as
            \be C_{\boldmath Z_2}:=\frac{C_{\uparrow}-C_{\downarrow}}{2}.\ee
        
        An example will be given to demonstrate that non zero $\boldmath Z_2$ truly indicates non-trivial physics. In fact, non zero $\boldmath Z_2$ charge is closely related to spin Hall effect and many other physical phenomena. As a typical meterial with non zero $\boldmath Z_2$ charge, $\rm Na_3Bi$ is a Dirac semimetal system, so the total Chern number is zero. Under low energy circumstance we can use following Hamiltonian $H(k)$ to describe the target system
            \be H=v_F(k_1\sigma^3\otimes\sigma^1-k_2\mathbb{I}_2\otimes\sigma^2)+m(k)\mathbb{I}_2\otimes \sigma^3,\ee
            in which the ``mass" term has the form $m(k)=-m_0+m_1{k_3}^2$ at low energy and $v_F$ denotes the Fermi velocity. This Hamiltonian can commute with spin operator $\sigma^3\otimes\mathbb{I}_2$, so the Hamiltonian can be separated into two blocks, each describing a Weyl semimetal with a single pair of Weyl nodes $H_\pm=v_F(\pm k_1\sigma^1-k_2\sigma^2)+m(k)\sigma^3$. After a series of calculation it is true that the system has non zero $\boldmath Z_2$ charge.
            
            In order to make it more clear how the non zero $\boldmath Z_2$ charge has an effect on the physics of the system, we will observe the contribution of non zero $\boldmath Z_2$ charge to charge current $j$ and spin current $\widetilde{j}$. We can write the charge current $j$ explicitly\cite{Burkov_2016}.
            \be j=\frac{e^2}{\pi^2}\mu_5 B+\widetilde{\sigma}_{xy}(\hat{z}\times \widetilde{E}).\ee
            
            The first term on the right hand side describes the chiral magnetic effect, i.e. a contribution to the charge current. The second term describes the inverse spin Hall effect, i.e. the generation of a charge current by a gradient of the spin density. Analogously we can write the spin current $\widetilde{j}$,
            \be \widetilde{j}=\frac{e^2}{\pi^2}\widetilde{\mu}_5 B+\widetilde{\sigma}_{xy}(\hat{z}\times E).\ee
            
            On the right hand, the first term describes the $\boldmath Z_2$ magnetic effect, which generates a contribution to the spin current, and the second term describe the spin Hall effect. $\widetilde{\sigma}_{xy}$ is the corresponding spin Hall conductivity, $\mu_5$ is the chiral chemical potential and $\widetilde{\mu}_5$ is the $\boldmath Z_2$ chemical potential. Only when there are non-zero $\boldmath Z_2$ charge, does the spin current and spin Hall effect exist.
        
         As demonstrated in the preceding example, it can be concluded that the $\boldmath Z_2$ charge is indeed capable of serving as the topological invariant for the spin degree of freedom. In the next part, we will calculate the $\boldmath Z_2$ charge in the holographic model  \eqref{holoz2}.
    \subsection{Topological invariant in the holographic Weyl-$\boldmath{Z}_2$ semimetal}
        As previously stated, two pairs of probe fermions are engaged in the detection of the chiral and $\boldmath Z_2$ charge within the holographic model \eqref{holoz2}, respectively. More narrowly, we use $\psi_1,\psi_2$ to detect the chiral charge and $\psi_3,\psi_4$ to detect the $\boldmath{Z_2}$ charge, i.e. each node has two topological invariant: the Weyl charge and the $\boldmath Z_2$ charge. Considering that $\psi_1$ only has coupling with $\psi_2$ and $\psi_3$ only has coupling with $\psi_4$, it is convenient to calculate the solution of $\psi_1,\psi_2$ and the solution of $\psi_3,\psi_4$ respectively. The topological Hamiltonian $H_{Weyl}$ is constructed by the boundary behaviour of $\psi_1,\psi_2$ and $H_{\boldmath{Z_2}}$ is constructed by the boundary behaviour of $\psi_3,\psi_4$, so we can first get $H_{Weyl}$ to calculate the Weyl charge for the nodes and then get $H_{\boldmath Z_2}$ to calculate the $\boldmath Z_2$ charge for the nodes.
        
Specifically, after obtaining $G_{chirality}$ and $G_{\boldmath Z_2}$ numerically for specific $k$, it is possible to get the topological Hamiltonian $H_{Weyl}(k)=-{G_{chirality}}^{-1}(0,k)$, $H_{\boldmath Z_2}(k)=-{G_{\boldmath Z_2}}^{-1}(0,k)$. To get the Weyl charge and $\boldmath Z_2$ charge, we can use the same procedure as in weakly coupled systems, i.e.
\begin{enumerate}
    \item Solve the eigen equations $H_{Weyl}(k)\ket{n(k)}_{Weyl}=E_n\ket{n(k)}_{Weyl}$, $H_{\boldmath Z_2}(k)\ket{n(k)}_{\boldmath Z_2}=E_n\ket{n(k)}_{\boldmath Z_2}$ to get the eigen states $\ket{n(k)}_{Weyl}, \ket{n(k)}_{\boldmath Z_2}$.
    \item Use the following formula to get the Chern class for each topological Hamiltonian.
        \begin{equation}
            c_{Weyl}=-\sum_{occ.}\frac{d\bra{n(k)}_{Weyl}\wedge d\ket{n(k)}_{Weyl}}{2\pi},~
            c_{\boldmath Z_2}=-\sum_{occ.}\frac{d\bra{n(k)}_{\boldmath Z_2}\wedge d\ket{n(k)}_{\boldmath Z_2}}{2\pi}.
        \end{equation}
    \item Choose the integral surfaces $\Sigma_{Weyl}$ and $\Sigma_{\boldmath Z_2}$ which enclose the Weyl node and the $\boldmath Z_2$ node respectively, and integrate the Chern class $c_{Weyl}$ and $c_{\boldmath Z_2}$ on the integral surfaces to get the corresponding Weyl charge and $\boldmath Z_2$ charge
        \begin{equation}
            Q_{Weyl}=\int_{\Sigma_{Weyl}}c_{Weyl},~
            Q_{\boldmath Z_2}=\int_{\Sigma_{\boldmath Z_2}}c_{\boldmath Z_2}.
        \end{equation}
\end{enumerate}
        \subsubsection{The chiral charge}
        We will first calculate the chiral charge and then afterwards we will show the calculation of $\boldmath Z_2$ charge, without loss of generality we choose $\frac{M_1}{b}=\frac{M_2}{c}=0.0477$.  As demonstrated in \eqref{z2probe}, the near horizon solution of the probe fermions presents eight free parameters, denoted by  $a_{+},a_{-}$. These parameters require eight independent initial (in-falling) conditions. In the presence of appropriate initial conditions, the coupled Dirac equation \eqref{z2femionequation} can be numerically resolved. The coefficients ${a_1}^i, {a_2}^i(i=1,2,...,8)$ in \eqref{z2uvc} represent the UV solution and can be utilised to determine both the source and the response matrix. The corresponding source and response matrices are  
        \be
        M_s=\begin{pmatrix}
            {a_1}^{1,\mathbf{I}} & {a_1}^{1,\mathbf{II}} & {a_1}^{1,\mathbf{III}} & {a_1}^{1,\mathbf{IV}} & {a_1}^{1,\mathbf{V}} & {a_1}^{1,\mathbf{VI}} & {a_1}^{1,\mathbf{VII}} & {a_1}^{1,\mathbf{VIII}} \\
            {a_1}^{2,\mathbf{I}} & {a_1}^{2,\mathbf{II}} & {a_1}^{2,\mathbf{III}} & {a_1}^{2,\mathbf{IV}} & {a_1}^{2,\mathbf{V}} & {a_1}^{2,\mathbf{VI}} & {a_1}^{2,\mathbf{VII}} & {a_1}^{2,\mathbf{VIII}} \\
            {a_1}^{3,\mathbf{I}} & {a_1}^{3,\mathbf{II}} & {a_1}^{3,\mathbf{III}} & {a_1}^{3,\mathbf{IV}} & {a_1}^{3,\mathbf{V}} & {a_1}^{3,\mathbf{VI}} & {a_1}^{3,\mathbf{VII}} & {a_1}^{3,\mathbf{VIII}} \\
            {a_1}^{4,\mathbf{I}} & {a_1}^{4,\mathbf{II}} & {a_1}^{4,\mathbf{III}} & {a_1}^{4,\mathbf{IV}} & {a_1}^{4,\mathbf{V}} & {a_1}^{4,\mathbf{VI}} & {a_1}^{4,\mathbf{VII}} & {a_1}^{4,\mathbf{VIII}} \\
            {a_2}^{5,\mathbf{I}} & {a_2}^{5,\mathbf{II}} & {a_2}^{5,\mathbf{III}} & {a_2}^{5,\mathbf{IV}} & {a_2}^{5,\mathbf{V}} & {a_2}^{5,\mathbf{VI}} & {a_2}^{5,\mathbf{VII}} & {a_2}^{5,\mathbf{VIII}} \\
            {a_2}^{6,\mathbf{I}} & {a_2}^{6,\mathbf{II}} & {a_2}^{6,\mathbf{III}} & {a_2}^{6,\mathbf{IV}} & {a_2}^{6,\mathbf{V}} & {a_2}^{6,\mathbf{VI}} & {a_2}^{6,\mathbf{VII}} & {a_2}^{6,\mathbf{VIII}} \\
            {a_2}^{7,\mathbf{I}} & {a_2}^{7,\mathbf{II}} & {a_2}^{7,\mathbf{III}} & {a_2}^{7,\mathbf{IV}} & {a_2}^{7,\mathbf{V}} & {a_2}^{7,\mathbf{VI}} & {a_2}^{7,\mathbf{VII}} & {a_2}^{7,\mathbf{VIII}} \\
            {a_2}^{8,\mathbf{I}} & {a_2}^{8,\mathbf{II}} & {a_2}^{8,\mathbf{III}} & {a_2}^{8,\mathbf{IV}} & {a_2}^{8,\mathbf{V}} & {a_2}^{8,\mathbf{VI}} & {a_2}^{8,\mathbf{VII}} & {a_2}^{8,\mathbf{VIII}}
        \end{pmatrix},
        \ee
        \be
        M_r=\begin{pmatrix}
            -{a_2}^{1,\mathbf{I}} & -{a_2}^{1,\mathbf{II}} & -{a_2}^{1,\mathbf{III}} & -{a_2}^{1,\mathbf{IV}} & -{a_2}^{1,\mathbf{V}} & -{a_2}^{1,\mathbf{VI}} & -{a_2}^{1,\mathbf{VII}} & -{a_2}^{1,\mathbf{VIII}} \\
            -{a_2}^{2,\mathbf{I}} & -{a_2}^{2,\mathbf{II}} & -{a_2}^{2,\mathbf{III}} & -{a_2}^{2,\mathbf{IV}} & -{a_2}^{2,\mathbf{V}} & -{a_2}^{2,\mathbf{VI}} & -{a_2}^{2,\mathbf{VII}} & -{a_2}^{2,\mathbf{VIII}} \\
            -{a_2}^{3,\mathbf{I}} & -{a_2}^{3,\mathbf{II}} & -{a_2}^{3,\mathbf{III}} & -{a_2}^{3,\mathbf{IV}} & -{a_2}^{3,\mathbf{V}} & -{a_2}^{3,\mathbf{VI}} & -{a_2}^{3,\mathbf{VII}} & -{a_2}^{3,\mathbf{VIII}} \\
            -{a_2}^{4,\mathbf{I}} & -{a_2}^{4,\mathbf{II}} & -{a_2}^{4,\mathbf{III}} & -{a_2}^{4,\mathbf{IV}} & -{a_2}^{4,\mathbf{V}} & -{a_2}^{4,\mathbf{VI}} & -{a_2}^{4,\mathbf{VII}} & -{a_2}^{4,\mathbf{VIII}} \\
            {a_1}^{5,\mathbf{I}} & {a_1}^{5,\mathbf{II}} & {a_1}^{5,\mathbf{III}} & {a_1}^{5,\mathbf{IV}} & {a_1}^{5,\mathbf{V}} & {a_1}^{5,\mathbf{VI}} & {a_1}^{5,\mathbf{VII}} & {a_1}^{5,\mathbf{VIII}} \\
            {a_1}^{6,\mathbf{I}} & {a_1}^{6,\mathbf{II}} & {a_1}^{6,\mathbf{III}} & {a_1}^{6,\mathbf{IV}} & {a_1}^{6,\mathbf{V}} & {a_1}^{6,\mathbf{VI}} & {a_1}^{6,\mathbf{VII}} & {a_1}^{6,\mathbf{VIII}} \\
            {a_1}^{7,\mathbf{I}} & {a_1}^{7,\mathbf{II}} & {a_1}^{7,\mathbf{III}} & {a_1}^{7,\mathbf{IV}} & {a_1}^{7,\mathbf{V}} & {a_1}^{7,\mathbf{VI}} & {a_1}^{7,\mathbf{VII}} & {a_1}^{7,\mathbf{VIII}} \\
            {a_1}^{8,\mathbf{I}} & {a_1}^{8,\mathbf{II}} & {a_1}^{8,\mathbf{III}} & {a_1}^{8,\mathbf{IV}} & {a_1}^{8,\mathbf{V}} & {a_1}^{8,\mathbf{VI}} & {a_1}^{8,\mathbf{VII}} & {a_1}^{8,\mathbf{VIII}}
        \end{pmatrix}.
        \ee
        
        The Green's function can thus be obtained as $G_{chirality}=i\Gamma^t M_r {M_s}^{-1}$, and the corresponding topological Hamiltonian can be determined accordingly, given by $H_{Weyl}=-{G_{chirality}}^{-1}$. $H_{Weyl}$ can be utilised to calculate the Chern number, i.e. the chiral charge, in the manner previously employed in the weak coupled system.
        
        \begin{figure}[ht]
            \centering
            \includegraphics[width=\textwidth]{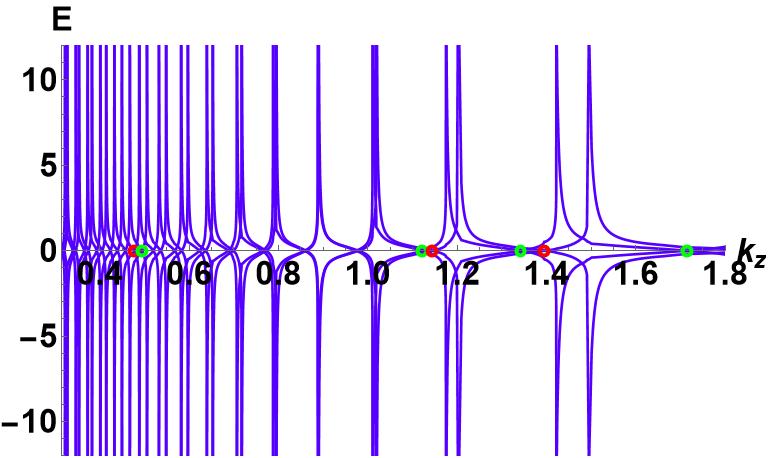}
            \caption{The Weyl charge for the nodes in the eigenvalue of $H_{Weyl}$ in a holographic Weyl $\boldmath Z_2$ semimetal, with  $\frac{M_1}{b}=0.0477$.  The red circle indicates a Weyl charge of 1 and the green circle -1. The nearest pair of nodes denoted with red and green circles is a pair of nodes with opposite topological charges, which will annihilate into a Dirac node.}
            \label{fig6}
        \end{figure} 
        
        Recall from Fig. \ref{fig3}, we can claim that the nodes for $H_{Weyl}$ only appear on the $k_z$ axis and the nodes for $H_{\boldmath Z_2}$ only appear on the $k_y$ axis. The Chern numbers of several pairs of Weyl nodes have been calculated, and the results demonstrate that each of the Weyl nodes carries the Chern number either $+1$ or $-1$, as is also the case in the weak coupled regime. The explicit values of topological invariants of several nodes calculated from numerics are illustrated in Fig. \ref{fig6}. The nearest nodes denoted with red and green circles have opposite chiral charges and form a pair of nodes which could annihilate in the critical phase transition point to a Dirac semimetal.  This has also been confirmed by the behavior of the band structure in the previous section. 
        
        Consequently, the numerical outcomes of the topological number demonstrate that when increasing $k_z$ from zero, the distances between a particular pair of Weyl nodes initially decrease to (almost) zero and subsequently increase. Without loss of generality, in Fig. \ref{fig6}, the Chern number of the three pairs of nodes around the values of $k_z$ equal to 0.48, 1.13 and 1.36, respectively, have been marked (red for $+1$ and green for $-1$). The results demonstrate that, for small $k_z$, in each pair of Weyl nodes, the node that has the negative Chern number sits on the left. Conversely, as $k_z$ increases, the node that has the positive Chern number is located on the left in place of the one with negative Chern number.
        
        As the system has multiple nodes, the red species of nodes are located in the axis repeatedly and so are the green ones. The distances between the red ones and the distances between the green ones are different so that there is a $k_{zc}\sim 0.76$, at which the red nodes transition from being positioned to the left of the green nodes to being situated to their right. This results in the distance between the paired nodes decreasing with $k_z$ at $k<k_{zc}$ and increasing at  $k>k_{zc}$.    It can thus be posited that the phenomenon of the order of the Chern number changing from $(-1, +1, -1...)$ to $(+1, -1, +1...)$ can be attributed to an interchange between two sets of bands resulting from the competition between two distinct scales: the scale dictating the distance between each red-green node pair and the scale determining the separation among the red nodes themselves. The first scale is determined by $\frac{M}{b}-(\frac{M}{b})_c$ and the second scale is determined by the conformal dimension and the charge of the probe fermion. 
        This is a very interesting behavior that is specially found in holographic systems, which we hope to also find in real condensed matter systems.

        \subsubsection{The  $\boldmath{Z}_2$ charge}

       As is the case in the chiral charge, the $\boldmath{Z}_2$ charge, which is detected by $\psi_3$ and $\psi_4$, can also be evaluated using the same procedure. The following matrices are provided for the source, denoted by $\hat{M}_s$, and the response, denoted by $\hat{M}_r$:
        \be
        \hat{M}_s=\begin{pmatrix}
            {a_3}^{1,\mathbf{I}} & {a_3}^{1,\mathbf{II}} & {a_3}^{1,\mathbf{III}} & {a_3}^{1,\mathbf{IV}} & {a_3}^{1,\mathbf{V}} & {a_3}^{1,\mathbf{VI}} & {a_3}^{1,\mathbf{VII}} & {a_3}^{1,\mathbf{VIII}} \\
            {a_4}^{2,\mathbf{I}} & {a_4}^{2,\mathbf{II}} & {a_4}^{2,\mathbf{III}} & {a_4}^{2,\mathbf{IV}} & {a_4}^{2,\mathbf{V}} & {a_4}^{2,\mathbf{VI}} & {a_4}^{2,\mathbf{VII}} & {a_4}^{2,\mathbf{VIII}} \\
            {a_3}^{3,\mathbf{I}} & {a_3}^{3,\mathbf{II}} & {a_3}^{3,\mathbf{III}} & {a_3}^{3,\mathbf{IV}} & {a_3}^{3,\mathbf{V}} & {a_3}^{3,\mathbf{VI}} & {a_3}^{3,\mathbf{VII}} & {a_3}^{3,\mathbf{VIII}} \\
            {a_4}^{4,\mathbf{I}} & {a_4}^{4,\mathbf{II}} & {a_4}^{4,\mathbf{III}} & {a_4}^{4,\mathbf{IV}} & {a_4}^{4,\mathbf{V}} & {a_4}^{4,\mathbf{VI}} & {a_4}^{4,\mathbf{VII}} & {a_4}^{4,\mathbf{VIII}} \\
            {a_4}^{5,\mathbf{I}} & {a_4}^{5,\mathbf{II}} & {a_4}^{5,\mathbf{III}} & {a_4}^{5,\mathbf{IV}} & {a_4}^{5,\mathbf{V}} & {a_4}^{5,\mathbf{VI}} & {a_4}^{5,\mathbf{VII}} & {a_4}^{5,\mathbf{VIII}} \\
            {a_3}^{6,\mathbf{I}} & {a_3}^{6,\mathbf{II}} & {a_3}^{6,\mathbf{III}} & {a_3}^{6,\mathbf{IV}} & {a_3}^{6,\mathbf{V}} & {a_3}^{6,\mathbf{VI}} & {a_3}^{6,\mathbf{VII}} & {a_3}^{6,\mathbf{VIII}} \\
            {a_4}^{7,\mathbf{I}} & {a_4}^{7,\mathbf{II}} & {a_4}^{7,\mathbf{III}} & {a_4}^{7,\mathbf{IV}} & {a_4}^{7,\mathbf{V}} & {a_4}^{7,\mathbf{VI}} & {a_4}^{7,\mathbf{VII}} & {a_4}^{7,\mathbf{VIII}} \\
            {a_3}^{8,\mathbf{I}} & {a_3}^{8,\mathbf{II}} & {a_3}^{8,\mathbf{III}} & {a_3}^{8,\mathbf{IV}} & {a_3}^{8,\mathbf{V}} & {a_3}^{8,\mathbf{VI}} & {a_3}^{8,\mathbf{VII}} & {a_3}^{8,\mathbf{VIII}}
        \end{pmatrix},
        \ee
        \be
        \hat{M}_r=\begin{pmatrix}
            -{a_4}^{1,\mathbf{I}} & -{a_4}^{1,\mathbf{II}} & -{a_4}^{1,\mathbf{III}} & -{a_4}^{1,\mathbf{IV}} & -{a_4}^{1,\mathbf{V}} & -{a_4}^{1,\mathbf{VI}} & -{a_4}^{1,\mathbf{VII}} & -{a_4}^{1,\mathbf{VIII}} \\
            {a_3}^{2,\mathbf{I}} & {a_3}^{2,\mathbf{II}} & {a_3}^{2,\mathbf{III}} & {a_3}^{2,\mathbf{IV}} & {a_3}^{2,\mathbf{V}} & {a_3}^{2,\mathbf{VI}} & {a_3}^{2,\mathbf{VII}} & {a_3}^{2,\mathbf{VIII}} \\
            -{a_4}^{3,\mathbf{I}} & -{a_4}^{3,\mathbf{II}} & -{a_4}^{3,\mathbf{III}} & -{a_4}^{3,\mathbf{IV}} & -{a_4}^{3,\mathbf{V}} & -{a_4}^{3,\mathbf{VI}} & -{a_4}^{3,\mathbf{VII}} & -{a_4}^{3,\mathbf{VIII}} \\
            {a_3}^{4,\mathbf{I}} & {a_3}^{4,\mathbf{II}} & {a_3}^{4,\mathbf{III}} & {a_3}^{4,\mathbf{IV}} & {a_3}^{4,\mathbf{V}} & {a_3}^{4,\mathbf{VI}} & {a_3}^{4,\mathbf{VII}} & {a_3}^{4,\mathbf{VIII}} \\
            {a_3}^{5,\mathbf{I}} & {a_3}^{5,\mathbf{II}} & {a_3}^{5,\mathbf{III}} & {a_3}^{5,\mathbf{IV}} & {a_3}^{5,\mathbf{V}} & {a_3}^{5,\mathbf{VI}} & {a_3}^{5,\mathbf{VII}} & {a_3}^{5,\mathbf{VIII}} \\
            -{a_4}^{6,\mathbf{I}} & -{a_4}^{6,\mathbf{II}} & -{a_4}^{6,\mathbf{III}} & -{a_4}^{6,\mathbf{IV}} & -{a_4}^{6,\mathbf{V}} & -{a_4}^{6,\mathbf{VI}} & -{a_4}^{6,\mathbf{VII}} & -{a_4}^{6,\mathbf{VIII}} \\
            {a_3}^{7,\mathbf{I}} & {a_3}^{7,\mathbf{II}} & {a_3}^{7,\mathbf{III}} & {a_3}^{7,\mathbf{IV}} & {a_3}^{7,\mathbf{V}} & {a_3}^{7,\mathbf{VI}} & {a_3}^{7,\mathbf{VII}} & {a_3}^{7,\mathbf{VIII}} \\
            -{a_4}^{8,\mathbf{I}} & -{a_4}^{8,\mathbf{II}} & -{a_4}^{8,\mathbf{III}} & -{a_4}^{8,\mathbf{IV}} & -{a_4}^{8,\mathbf{V}} & -{a_4}^{8,\mathbf{VI}} & -{a_4}^{8,\mathbf{VII}} & -{a_4}^{8,\mathbf{VIII}}
        \end{pmatrix}.
        \ee
        
        \begin{figure}[h!]
            \centering
            \includegraphics[width=\textwidth]{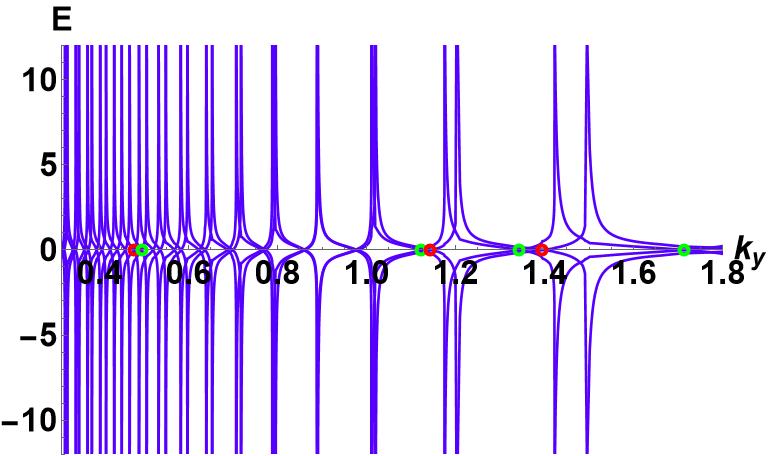}
            \caption{The $\boldmath Z_2$ charge for the nodes in the eigenvalue of $H_{\boldmath Z_2}$ in a holographic Weyl $\boldmath Z_2$ semimetal, with  $\frac{M_2}{c}=0.0477$.  The red circle indicates a $\boldmath Z_2$ charge of 1 and the green circle -1. The nearest pair of nodes denoted with red and green circles is a pair of nodes with opposite topological charges, which will annihilate into a Dirac node.}
            \label{fig7}
        \end{figure}
        
        The Green's function, which corresponds to the $\boldmath{Z}_2$ degree of freedom, is given by the following equation: $G_{\boldmath Z_2}=i\hat{\Gamma}^t \hat{M}_r {\hat{M}_s}^{-1}$. The effective Hamiltonian $H_{\boldmath Z_2}=-{G_{\boldmath{Z}_2}}^{-1}$ can be obtained directly.
         
        In Fig. \ref{fig7}, the energy dispersion, determined by $H_{\mathbb{Z}_2}$, is plotted as a function of $k_y$. The value of the $\boldmath{Z}_2$ charge for the three pairs of nodes has also been marked, which also equal to $+1$(read points) or $-1$(green points). In a manner analogous to the phenomenon observed in the chiral degree of freedom, the energy bands undergo an exchange following the attainment of a critical value of $k_{yc}$. The chiral charge is found to be zero for nodes located along the $k_y$ axis, while the $\boldmath{Z}_2$ also vanish when the nodes located along the $k_z$ axis.

        In summary, the topological charge for the nodes in the holographic model \eqref{holoz2} can be expressed as follows: $(1,0),~(1,-0),~(0,1),~(0,-1)$. Within the braket, the first is identified as the chiral charge, and the second as the $\boldmath Z_2$ charge. With regard to the configuration depicted on the right side of Fig. \ref{fig3}, the topological charge would be designated as $(1,1),~(1,-1),~(-1,1),~(-1,-1)$. This suggests an alternative topological structure, which will not be further discussed in this paper.

%% file: Section/Section5.tex
\section{Discussion and Outlook}
\label{sec:5}
    The topological invariant serves as the fundamental symbol of the distinct topological phases. In this paper, we have calculated the chiral charge and $\boldmath Z_2$ charge, which can be treated as the topological invariant of the Weyl/$\boldmath{Z_2}$ semimetal in holographic model. By employing four fermions to probe the holographic Weyl/$\boldmath Z_2$ semimetal and utilising the topological Hamiltonian approach, the Weyl and $\boldmath Z_2$ charges of each Weyl point have been accurately fixed. The Weyl and $\boldmath Z_2$ charges, in combination, comprise the corresponding topological invariant. The findings reveal that within the holographic model \eqref{holoz2}, the corresponding topological invariant of the four Weyl nodes are $(1,0), (0,1), (-1,0)$ and $(0,-1)$, respectively.

    Furthermore, the existence of multi-Fermi surface structures has been observed in the holographic model, as indicated by calculations of fermion spectrum. The emergence of these multi-Fermi surfaces can be attributed to the phenomenon of strong coupling in holographic model. In turn, the model \eqref{holoz2} contains multiple pairs of Weyl/$\boldmath Z_2$ nodes. We have confirmed which two adjacent Weyl/$\boldmath Z_2$ nodes should be a pair of nodes with opposite topological charges that could be annihilated to form a Dirac node at the topological phase transition critical point, both from topological charges and the analysis of the band structures when tuning parameters of the system that determine the distance between each pair of nodes.  
    It is important to highlight that these phenomena are not readily discernible through a mere examination of the $\sigma_{AHE}$ alone. However, further investigation through ARPES may prove beneficial in elucidating this phenomenon.
    
    From another perspective, the Weyl/$\boldmath{Z}_2$ semimetal can be conceptualised as a coexistence state between two ideal Weyl semimetals. However, the Weyl/$\boldmath{Z}_2$ semimetal exhibits markedly distinct topological properties compared to the ideal Weyl semimetal. Therefore, in order to gain a deeper understanding of the topological phase transition, it is imperative to consider the many-body system which realises the coexistence of disparate topological phases. It has been proposed that a different type of topological many-body system, the topological nodal line semimetal, is capable of coexisting with an ideal Weyl semimetal. Furthermore, a holographic model that realises such a coexisting state has been suggested. In subsequent research, the intention is to calculate the corresponding topological number of the coexisting state, as well as to examine the energy band exchange phenomenon\cite{future}. 
    
    Finally, it should be noted that this paper defines the topological Hamiltonian based on the zero-temperature Green's function. It is essential to investigate the impact of finite temperature on the stability of the topological structure, as has been done in \cite{Song:2019asj}. Furthermore, the presence of impurities in a real many-body system is inevitable, and this has a consequential effect on the transport behaviour of the system. For example, the magneto-conductivity of the weak-coupling Weyl semimetal exhibits distinct behaviour in response to different impurities\cite{im1,im2}. It is crucial to investigate the effect of these impurities under the strong coupling limit on the topological structure and the corresponding topological phase transitions. 
    

%% file: Appendices.tex
\appendix
\section{Spin connection in curved space}
    \label{AppendixA}
    Given a manifold $M$ equipped with a metric $g$, we may choose an orthonormal frame $\{e_i\}$ and its dual coframe $\{\theta^i\}$ without loss of generality. The connection 1-form ${\omega_j}^i$ defines the covariant derivative as
    \be
        D e_i = {\omega_i}^j \otimes e_j.
    \ee

    The corresponding connection form $\omega$ on the frame bundle $F(M)$ relates to the connection 1-form on the base manifold $M$ via the pullback by a section $\sigma: M \to F(M)$:
    \be
        \sigma^*\omega := \begin{pmatrix}
            {\omega_1}^1 & \cdots & {\omega_n}^1 \\
            \vdots & \ddots & \vdots \\
            {\omega_1}^n & \cdots & {\omega_n}^n
        \end{pmatrix}: T_p M \to \mathfrak{o}(n),
    \ee
    where $\sigma(p) = (p, e_1, \dots, e_n)$. If the Stiefel-Whitney classes $w_1(M)$ and $w_2(M)$ vanish, the frame bundle $F(M)$ reduces to a spin structure $Sp(M)$ with structure group $Spin(n)$ and homomorphism $\rho: Spin(n) \to SO(n)$.

    The local connection form $\hat{\omega}$ on $Sp(M)$ is induced by $\omega$ via $\hat{\omega} = (\rho_*)^{-1} \circ \sigma^* \omega$. For the Levi-Civita connection (with ${\omega_j}^i = -{\omega_i}^j$), this simplifies to
    \be
        \hat{\omega} = \frac{1}{2} \sum_{i<j} {\omega_i}^j e_i e_j: T_p M \to \frak{spin}(n),
    \ee
    where $e_i e_j \in Cl(T_p M)$ is the Clifford product. The spinor bundle is built from the $Spin(n)$-representation $\tau:Spin(n)\to GL(\Psi)$. Let $Sp(M)$ be a spin structure on $M$. Using the representation $\tau$, the associated vector bundle $S(M):=Sp(M)\times_\tau\Psi$ is defined by quotienting $Sp(M)\times\Psi$ under the equivalence relation $(p\cdot g,v)\sim(p,\tau(g)v)$ for $g\in Spin(n)$. $S(M)$ is a vector bundle with fiber $\Psi$, whose sections are spinor fields.
        
    The covariant derivative of a spinor field $\Psi$ is given by\cite{lawson2016spin}:
    \begin{equation}
        D\Psi=d\Psi+\hat{\omega}\cdot\psi=d\Psi+\frac{1}{4}\sum_{ij} {\omega_i}^j (e_i\cdot e_j-e_j\cdot e_i)\cdot\Psi,
    \end{equation}
    where $\cdot$ denotes the Clifford product. In terms of gamma matrices, this can be rewritten in a more conventional form by defining $\Gamma^a:=\eta^{ab}e_b\cdot$ in the Lorentzian manifold:
    \begin{equation}
        D\Psi=d\Psi+\frac{1}{4}\sum_{ab} {\omega_a}^b[\Gamma^a,\Gamma^b]\Psi.
    \end{equation}
    
    Here, $\eta = {rm diag\ }(1,-1,-1,-1,-1)$ is the metric signature, and the spinor space inner product is defined as $(\Psi_1,\Psi_2):=i\Psi_1^\dag\Gamma^0\Psi_2$. This formulation satisfies the Clifford algebra relations:
    \begin{equation}
        {\Gamma^a,\Gamma^b}=-2\eta^{ab}, \quad (\Psi_1,\Gamma^a\Psi_2)=-(\Gamma^a\Psi_1,\Psi_2)
    \end{equation}
    
    The Dirac operator is given by $\Gamma^a D_{e_a}$, which leads to the equation of motion for a free fermion of mass $m\in\mathbb{R}$:
    \begin{equation}
        \Gamma^aD_{e_a}\Psi-m\Psi=0.
    \end{equation}
        
    In particular, when the metric takes the simple diagonal form
    \begin{equation}
        g = \sum_a \eta^{aa} g_{aa} dx^a \otimes dx^a,
    \end{equation}
    the connection 1-form can be expressed explicitly. First, we define the orthonormal frame and its dual coframe:
    \begin{equation}
        \theta^a = \sqrt{g_{aa}} dx^a, \quad e_a = \frac{1}{\sqrt{g_{aa}}} \frac{\partial}{\partial x^a}.
    \end{equation}

    Using Cartan's structure equation
    \begin{equation}
        d\theta^a = \theta^b \wedge \omega_b{}^a,
    \end{equation}
    we obtain the explicit expression for the connection 1-form $\omega_b{}^a$.
         
    \be
        {\omega_b}^a=\frac{1}{\sqrt{g_{bb}}}\frac{\partial\sqrt{g_{aa}}}{\partial x^b}dx^a+\frac{1}{\sqrt{g_{aa}}}\frac{\partial\sqrt{g_{bb}}}{\partial x^a}dx^b,~~a=0,b\ne 0 ~or~a\ne 0,b=0,\nonumber
    \ee
    \be
        {\omega_b}^a=\frac{1}{\sqrt{g_{bb}}}\frac{\partial\sqrt{g_{aa}}}{\partial x^b}dx^a-\frac{1}{\sqrt{g_{aa}}}\frac{\partial\sqrt{g_{bb}}}{\partial x^a}dx^b,~~else.
    \ee

        The covariant derivative can be written explicitly
        \be
        D_{e_c}\Psi=\frac{1}{\sqrt{g_{cc}}}\frac{\partial\Psi}{\partial x^c}+\frac{1}{4}\sum_{ab}\braket{{\omega_a}^b,e_c}[\Gamma^a,\Gamma^b]\Psi.
        \ee
        
        Without loss of generality, we choose the Gamma matrices as chiral-Weyl representation.
        \be
        I_2=\begin{pmatrix}1 & 0 \\ 0 & 1\end{pmatrix},~\sigma^1=\begin{pmatrix}0 & 1 \\ 1 & 0\end{pmatrix},~\sigma^2=\begin{pmatrix}0 & -i \\ i & 0\end{pmatrix},~\sigma^3=\begin{pmatrix}1 & 0 \\ 0 & -1\end{pmatrix},\nonumber
        \ee
        \be
        \gamma^0=\begin{pmatrix}0 & I_2 \\ I_2 & 0\end{pmatrix},~\gamma^1=\begin{pmatrix}0 & \sigma^1 \\ -\sigma^1 & 0\end{pmatrix},~\gamma^2=\begin{pmatrix}0 & \sigma^2 \\ -\sigma^2 & 0\end{pmatrix},~\gamma^3=\begin{pmatrix}0 & \sigma^3 \\ -\sigma^3 & 0\end{pmatrix},\nonumber
        \ee
        \be
        \gamma^5=i\gamma^0\gamma^1\gamma^2\gamma^3=\begin{pmatrix}-I_2 & 0 \\ 0 & I_2 \end{pmatrix},\nonumber
        \ee
        \be
        (\Gamma^t,\Gamma^x,\Gamma^y,\Gamma^z,\Gamma^r)=(i\gamma^0,i\gamma^1,i\gamma^2,i\gamma^3,-\gamma^5).
        \ee
        
    In this article, we adopt the abbreviated notation:
    \begin{equation}
        D_a \equiv D_{e_a}
    \end{equation}
    for the covariant derivative along the frame field $e_a$.
\section{Chern class}
    The Chern class is a cohomology class in ${H_{\text{dR}}}^{2k}(M;\mathbb{R})$ that characterizes the topological structure of a complex vector bundle $(E,M,\pi)$. We begin with the fundamental concept of invariant polynomials.

    \begin{definition}[Invariant Polynomial]
        Let $V$ be a finite-dimensional vector space. For $k \geq 1$, consider a multilinear map:
        \begin{equation}
            f: \underbrace{V \times \cdots \times V}_{k \text{ times}} \to \mathbb{R}, \quad (v_1,\ldots,v_k) \mapsto f(v_1,\ldots,v_k)
        \end{equation}
        satisfying:
        \begin{enumerate}
            \item Linearity in each argument $v_i$,
            \item Symmetry under permutation: $f(v_{\sigma(1)},\ldots,v_{\sigma(k)}) = f(v_1,\ldots,v_k)$ for all $\sigma \in S_k$.
        \end{enumerate}
        The space of such maps forms a vector space $S^k(V^*)$, with $S^0(V^*) := \mathbb{R}$.

        For a Lie group $G$ with Lie algebra $\mathfrak{g}$, a polynomial $f \in S^k(\mathfrak{g}^*)$ is called \emph{invariant} if:
        \begin{equation}
            f(\text{Ad}_{g^{-1}}X_1,\ldots,\text{Ad}_{g^{-1}}X_k) = f(X_1,\ldots,X_k) \quad \forall g \in G
        \end{equation}

        The space of degree-$k$ invariant polynomials is denoted:
        \begin{equation}
            I^k(G) := { f \in S^k(\mathfrak{g}^*) : f \text{ is invariant} }
        \end{equation}
        with $I^*(G) := \bigoplus_{k\geq 0} I^k(G)$.

        For $V$-valued differential forms $\eta_1,\ldots,\eta_k$ on a manifold, expressed in a basis ${E_i}$ as $\eta_i = \eta_i^j E_j$, we extend the definition:
        \begin{equation}
            f(\eta_1,\ldots,\eta_k) := f(E_{j_1},\ldots,E_{j_k}) \eta_1^{j_1} \wedge \cdots \wedge \eta_k^{j_k}
        \end{equation}
    \end{definition}
    
    For our purposes, the properties of $I^k(GL(r,\mathbb{R}))$ are particularly significant.

    \begin{proposition}[Differential Properties of Invariant Polynomials]
        Let $f \in I^k(GL(r,\mathbb{R}))$ be an invariant polynomial. Then:
        \begin{enumerate}
            \item For all $X, X_i \in \mathfrak{gl}(r,\mathbb{R})$,
                \begin{equation}
                    \sum_{i=1}^k f(X_1, \ldots, [X_i, X], \ldots, X_k) = 0
                \end{equation}
            \item For $\mathfrak{gl}(r,\mathbb{R})$-valued differential forms $\eta, \eta_i$ of odd degree,
                \begin{equation}
                    \sum_{i=1}^k f(\eta_1, \ldots, [\eta_i, \eta], \ldots, \eta_k) = 0
                \end{equation}
        \end{enumerate}
    \end{proposition}

    \begin{proof}
        The invariance of $f$ implies that for any $t \in \mathbb{R}$ and $X \in \mathfrak{gl}(r,\mathbb{R})$:
        \begin{equation}
            f(e^{-tX}X_1e^{tX}, \ldots, e^{-tX}X_ke^{tX}) = f(X_1, \ldots, X_k)
        \end{equation}

        Differentiating at $t=0$ yields the first result. The second property follows by expanding $\eta$ and $\eta_i$ in a basis of $\mathfrak{gl}(r,\mathbb{R})$ and applying the first property.
    \end{proof}

    Now consider a vector bundle $(E,M,\pi)$ with structure group $GL(r,\mathbb{R})$, equipped with a connection $D$. Locally, the connection acts on a frame ${e_i}$ of $E$ as:
    \begin{equation}
        D e_i = {\omega_i}^j \otimes e_j
    \end{equation}
    where ${\omega_j{}^i}$ are the connection 1-forms. These can be organized into a matrix-valued 1-form:
    \begin{equation}
         \omega:=\begin{pmatrix}
            {\omega_1}^1 & {\omega_2}^1 & \cdots & {\omega_r}^1\\
            {\omega_1}^2 & {\omega_2}^2 & \cdots & {\omega_r}^2\\
            \vdots & \vdots & \ddots & \vdots\\
            {\omega_1}^r & {\omega_2}^r & \cdots & {\omega_r}^r
        \end{pmatrix},
    \end{equation}
    
    The connection form $\omega$ is a $\mathfrak{gl}(r,\mathbb{R})$-valued 1-form, whose curvature $\Omega$ is a $\mathfrak{gl}(r,\mathbb{R})$-valued 2-form defined by:
    \begin{equation}
        \Omega := d\omega + \frac{1}{2}[\omega,\omega].
    \end{equation}

    For any invariant polynomial $f \in I^k(GL(r,\mathbb{R}))$, we define:
    \begin{equation}
        f(\Omega) := \underbrace{f(\Omega,\ldots,\Omega)}_{k\text{ times}}.
    \end{equation}

    By the invariance property, $f(\Omega)$ is well-defined globally on the base manifold $M$. This construction has the following crucial properties:

    \begin{proposition}[Properties of Characteristic Forms]
        For any connection on $(E,M,\pi)$, the form $f(\Omega)$ satisfies:
        \begin{enumerate}
            \item It is closed: $f(\Omega) \in {\Omega_{cl}}^{2k}(M)$
            \item Its cohomology class $[f(\Omega)] \in {H_{\text{dR}}}^{2k}(M)$ is independent of the connection choice
        \end{enumerate}
    \end{proposition}

    \begin{proof}
        For closedness, we compute:
        \begin{align*}
            df(\Omega) &= \sum_{i=1}^k f(\Omega,\ldots,d\Omega,\ldots,\Omega) \\
            &= \sum_{i=1}^k f(\Omega,\ldots,[\Omega,\omega],\ldots,\Omega) \quad \text{(Bianchi identity)} \\
            &= 0 \quad \text{(by Proposition 2.2)}.
        \end{align*}

        For connection independence, consider an interpolation $\omega_t = (1-t)\omega_0 + t\omega_1$ between two connections, with $\eta = \omega_1 - \omega_0$. Then:
        \begin{align*}
            f(\Omega_1) - f(\Omega_0) &= \int_0^1 \frac{d}{dt}f(\Omega_t) dt \\
            &= k\int_0^1 f(d\eta + [\eta,\omega_t],\Omega_t,\ldots,\Omega_t) dt \\
            &= d\left(k\int_0^1 f(\eta,\Omega_t,\ldots,\Omega_t) dt\right) =: d\Phi.
        \end{align*}
    \end{proof}

    This construction gives rise to the \emph{Chern-Weil homomorphism}:
    \begin{equation}
        w(E,\cdot): I^*(G) \to {H_{\text{dR}}}^*(M), \quad f \mapsto [f(\Omega)].
    \end{equation}

    \begin{definition}[Chern Classes]
        For the unitary group $U(n) \subset GL(n,\mathbb{C})$ with Lie algebra $\mathfrak{u}(n) = {A \in \mathfrak{gl}(n,\mathbb{C}) | A = -A^\dagger}$, consider the characteristic polynomial:
        \begin{equation}
            \det\left(\lambda I_n - \frac{1}{2\pi i}A\right) = \sum_{k=0}^n c_k(A)\lambda^{n-k}.
        \end{equation}

        The coefficients $c_k$ satisfy:
        \begin{itemize}
            \item Homogeneity: $c_k(mA) = m^k c_k(A)$
            \item Ad-invariance: $c_k(gAg^{-1}) = c_k(A)$ for $g \in U(n)$
            \item Reality: $c_k(A) \in \mathbb{R}$ for $A \in \mathfrak{u}(n)$
        \end{itemize}

        The \emph{$k$-th Chern class} of $E$ is:
        \begin{equation}
            c_k(E) := [c_k(\Omega)] \in {H_{\text{dR}}}^{2k}(M;\mathbb{R})
        \end{equation}
        which is a topological invariant independent of the connection.
    \end{definition}
\section{Euler characteristic and Chern number}
    Following the definition of Chern classes, we now examine a concrete example in 2-dimensional systems where the Chern number directly relates to the Euler characteristic. This connection demonstrates how Chern classes capture fundamental topological properties of physical systems.

    Consider a $d$-dimensional quantum system with Hamiltonian $\hat{H}(k)$, where $k$ belongs to the Brillouin zone $\mathbb{T}^d$. For an eigenenergy branch $E_n(k)$ with corresponding eigenvectors $\ket{n(k)}$ satisfying
    \begin{equation}
        \hat{H}(k)\ket{n(k)} = E_n(k)\ket{n(k)},
    \end{equation}
    we observe that the phase freedom ($\ket{n(k)} \to e^{i\theta(k)}\ket{n(k)}$) allows us to construct a complex line bundle:
    \begin{equation}
        V_n := \bigsqcup_{k\in\mathbb{T}^d} \{\alpha\ket{n(k)} | \alpha \in \mathbb{C}\} .
    \end{equation}

    For this $U(1)$-bundle $V_n$, the first Chern class takes a particularly simple form. From the characteristic polynomial
    \begin{equation}
        \det\left(\lambda - \frac{A}{2\pi i}\right) = c_0(A)\lambda + c_1(A), \quad A \in \mathfrak{u}(1),
    \end{equation}
    we obtain the topological invariant:
    \begin{equation}
        c_1(V_n) = -\frac{[\Omega]}{2\pi i} \in H^2_{\text{dR}}(\mathbb{T}^d),
    \end{equation}
    where $\Omega$ is the curvature 2-form.

    The natural choice of connection (Berry connection) in this physical context is:
    \begin{equation}
        \omega := i\bra{n(k)}d\ket{n(k)} \in \Omega^1(\mathbb{T}^d,\mathfrak{u}(1)),
    \end{equation}
    with corresponding covariant derivative
    \begin{equation}
        D\ket{n(k)} = \omega \otimes \ket{n(k)}.
    \end{equation}
    
    The connection we defined satisfies all requirements of a principal $U(1)$-connection. Through Cartan's structure equation, we obtain the curvature form:
    \begin{equation}
        \Omega = d\omega - \omega \wedge \omega = i,d\bra{n(k)}\wedge d\ket{n(k)},
    \end{equation}
    which leads to the first Chern class for normalized eigenvectors:
    \begin{equation}
        c_1 = -\frac{[\Omega]}{2\pi i} = -\frac{[d\bra{n(k)}\wedge d\ket{n(k)}]}{2\pi} \in H^2(\mathbb{T}^d,\mathbb{R}).
    \end{equation}

    For 2D systems ($d=2$), we define the Chern number of the energy band $E_n$ as:
    \begin{equation}
        C(E_n) := \int_{\mathbb{T}^2} c_1 = -\frac{1}{2\pi}\int_{\mathbb{T}^2} d\bra{n(k)}\wedge d\ket{n(k)}.
    \end{equation}

    This integral is well-defined since exact forms integrate to zero by Stokes' theorem.

    The eigenspace
    \begin{equation}
        V_n(k) =  \{\alpha\ket{n(k)} | \alpha \in \mathbb{C}\}
    \end{equation}
    admits a natural realification:

    \begin{definition}[Realification]
        Let $(^\mathbb{C}V, (\cdot,\cdot)_\mathbb{C})$ be a complex vector space. Its realification $^\mathbb{R}V$ is the real vector space equipped with the inner product:
        \begin{equation}
            (X,Y)_\mathbb{R} := \Re(X,Y)_\mathbb{C}.
        \end{equation}

        This preserves norms through the polarization identities:
        \begin{align}
            (X,Y)_\mathbb{C} &= \frac{||X+Y||^2 - ||X-Y||^2}{4} + i\frac{||X+iY||^2 - ||X-iY||^2}{4}, \\
            (X,Y)_\mathbb{R} &= \frac{||X+Y||^2 - ||X-Y||^2}{4}.
        \end{align}
        The realification process effectively "forgets" the complex structure while maintaining the metric properties.
    \end{definition}

    Through realification, we obtain a real vector bundle $W_n$ with fibers:
    \begin{equation}
        W_n(k) = \text{span}_\mathbb{R}\{\ket{n(k)}, i\ket{n(k)}\},
    \end{equation}
    equipped with the real inner product:
    \begin{equation}
        (\alpha\ket{n(k)}, \beta\ket{n(k)}) = \Re(\overline{\alpha}\beta)\braket{n(k)|n(k)}.
    \end{equation}

    This construction naturally leads to a Riemannian interpretation:
    \begin{definition}[Quantum State Manifold]
        For a 2D system with Hamiltonian $\hat{H}(k)$ and eigenstate $\ket{n(k)}$ satisfying:
        \begin{equation}
            \hat{H}(k)\ket{n(k)} = E_n(k)\ket{n(k)}, \quad k \in \mathbb{T}^2,
        \end{equation}
        we define a closed surface $M_n$ with:
        \begin{itemize}
            \item Orthonormal frame: $e_1 = \ket{n(k)}$, $e_2 = i\ket{n(k)}$.
            \item Dual coframe: $\theta^1 = \Re\bra{n(k)}$, $\theta^2 = \Re(-i\bra{n(k)})$.
            \item Riemannian metric: $g = \theta^1 \otimes \theta^1 + \theta^2 \otimes. \theta^2$
        \end{itemize}
    \end{definition}

    The geometry of $M_n$ reveals deep connections:
    \begin{itemize}
        \item The Levi-Civita connection 1-form is:
            \begin{equation}
               \omega_1{}^2 = -\omega_2{}^1 = i\bra{n(k)}d\ket{n(k)},
            \end{equation}
            satisfying Cartan's structure equations $d\theta^i = \theta^j \wedge \omega_j{}^i$.

        \item The curvature 2-form becomes:
            \begin{equation}
                \Omega_2{}^1 = -d\bra{n(k)} \wedge d\ket{n(k)},
            \end{equation}
            where the inner product is taken in the realified space.
    \end{itemize}

    Applying the Gauss-Bonnet theorem yields:
    \begin{equation}
        \chi(M_n) = \frac{1}{2\pi}\int_{M_n} \Omega_2{}^1 = -\frac{1}{2\pi}\int_{\mathbb{T}^2} d\bra{n(k)} \wedge d\ket{n(k)} = C(E_n),
    \end{equation}
    establishing the exact equality between the Euler characteristic and Chern number.

    Key observations about degeneracies:
    \begin{itemize}
        \item At degeneracy points $k_0$ where $E_n(k_0)$ is not isolated, the eigenstate $\ket{n(k_0)}$ becomes ill-defined.
        \item The Poincaré-Hopf theorem requires that when $\chi(M_n) \neq 0$, the vector field must have zeros corresponding to band degeneracies.
        \item These degeneracies are topologically protected - they cannot be removed by small perturbations.
    \end{itemize}

    The Chern number's topological nature implies:
    \begin{itemize}
        \item Robustness against continuous deformations of the Hamiltonian.
        \item Quantized Hall conductivity in quantum Hall systems.
        \item Protection of edge states in topological insulators.
    \end{itemize}
\section{Concise explanation for effectiveness of topological Hamiltonian}
\label{AppendixD}
The topological Hamiltonian is defined in terms of the zero-frequency Green's function. It is therefore pertinent to inquire whether potential poles in the complex frequency domain impact the validity of the topological Hamiltonian method. In this appendix, we will  clarify why the topological Hamiltonian approach remains valid without requiring that the Green's function has no singularities or poles in the imaginary frequency space, and justify our use of real-frequency Green's functions for holographic semimetals.

As derived in the original work proposing the topological Hamiltonian approach  \cite{wang-prx1}, the topological invariant in terms of Green's functions for interacting systems could be calculated from
    \begin{equation}\label{d1}
        N_2=\frac{1}{24\pi ^2}\int d k_0 d^2k \text{Tr}[\epsilon^{\mu\nu\rho}G\partial_\mu G^{-1}G\partial_\nu G^{-1}G\partial_\rho G^{-1}],
    \end{equation}
    where $G(w,k)$ is the Green's functions. In the original work, the authors have used imaginary frequency Green’s functions, which are the Matsubara Green’s functions, while emphasizing that their final conclusions are also applicable using real frequency. The reason that they used imaginary frequency Green's functions is that this could be used at finite temperature and the imaginary frequency Green's functions are more well behaved in their case. At the same time calculations on real frequency Green's functions for interacting fermionic systems are much more complicated than imaginary frequency ones in condensed matter physics. 

   The topological Hamiltonian approach relies on a smooth deformation of the Green's function
    \begin{equation}
        G(iw,k,\lambda)=(1-\lambda)G(iw,k)+\lambda [i\omega+G^{-1}(0,k) ]^{-1},
    \end{equation} where $\lambda \in [0,1]$. They have shown that $G(iw,k,\lambda)$ is a smooth function so tuning from $\lambda=0$ to $\lambda=1$ does not modify the topological properties. Therefore, using the effective Green's function $[i\omega+G^{-1}(0,k)] $ (corresponding to $\lambda=1$) would give the original and correct topological invariant (corresponding to $\lambda=0$). Then substituting the $\lambda=1$ Green's function into the formula \eqref{d1}, the calculation of the topological invariant would reduce to the weakly coupled calculation using $G^{-1}(0,k)$ as an effective Hamiltonian. In the process to show that tuning $\lambda$ from zero to one is a smooth process, they have used the fact that $G(i\omega,k,\lambda)$ does not have zero eigenvalues at nonzero $\omega$, seemingly introducing a constraint that the Green's function has to have no poles in the imaginary frequency space other than $\omega=0$. 

    Let us then move to the calculation of the topological invariant for a semimetal. In our case, to calculate the topological invariant for a semimetal, which is a gapless topological state, the topological invariant should be always calculated in a surface/circle surrounding the node, where the node is a zero point of the Green's functions. Thus there is a small modification to that definition of the universal definition of the topological invariant, which now becomes
    \begin{equation}
        N=\frac{1}{24\pi ^2}\int d \omega \oint d^2k \text{Tr}[\epsilon^{\mu\nu\rho}G\partial_\mu G^{-1}G\partial_\nu G^{-1}G\partial_\rho G^{-1}].
    \end{equation} The integral on $k$ is restricted to the small surface enclosing the Weyl node in the momentum space while the integral in $\omega$ covers the whole real $\omega$ axis. This integration in the $k$ space automatically avoids the singular points or poles of the Green's functions as the enclosing surface could be chosen to be extremely close to the Weyl point. 

   Note that at the same time, as we have explained, in the original work for topological Hamiltonian, the authors introduced imaginary frequency Green's functions for their calculational convenience. In our case, it should be more straightforward and more justified to directly use the real frequency Green's function. The reasons are that: the real-frequency Green's function directly probes the physical spectral function, making it ideal for studying the low-energy physics of Weyl points and their topological properties; while the imaginary-frequency Green's function is less directly tied to physical observables, as it represents the system in a Euclidean framework. Analytic continuation to the real axis may be needed to extract physical properties, which introduces numerical errors.

    For strongly interacting Weyl semimetals, imaginary frequency Green's functions would be much easier to calculate though not so relevant for physical systems, however, holography solves this problem so that we could use directly the real frequency Green's functions. Therefore, in the formula for the calculation of topological invariants for holographic semimetals, the frequency should be only integrated in the real space. Thus as long as there are no poles/singlar points in real frequency space at $\omega \neq 0$, which is indeed true, the topological Hamiltonian approach is justified using the tuning $\lambda$ method. Especially, the integration in the $k$ space is performed over an extremely small enclosing surface around the Weyl point. This ensures that it is quite easy to verify that there are no singular points or poles for the Green's functions within the entire frequency range when k is restricted to be extremely close to the Weyl point.

    Thus, we could instead define the following real frequency version of the smooth function
    \begin{equation}
        G(w,k,\lambda)=(1-\lambda)G(w,k)+\lambda [\omega+G^{-1}(0,k) ]^{-1}.
    \end{equation} As long as $G(w,k,\lambda)$ is regular for nonzero $\omega$ in the real frequency space, the topological invariant calculated for $\lambda=1$ would be the same for $\lambda=0$, i.e. we could use the topological Hamiltonian method to use $G^{-1}(0,k)$ as the effective Hamiltonian for topological invariants. 

    Therefore, in our work, we do not require the Green's function to have no poles in the imaginary frequency space. For our use, the topological Hamiltonian method is justified and the topological invariants are robust and well-captured by the zero-frequency limit of the Green's function. This also reflects the fact that topology is encoded in the global properties of the Green's function, and the zero-frequency slice often retains this essential topological information.